\RequirePackage{ifpdf}
\documentclass{PoS}
\pdfoutput = 1
\usepackage{cite}
\usepackage{amsmath}  
\usepackage{amssymb} 
\usepackage{dsfont}
 \usepackage{multirow}
\usepackage{psfrag,scalefnt}
\usepackage{paralist}

\usepackage{color}
\let\normalcolor\relax

\newcommand{\vus}{|V_{us}|}
\newcommand{\vcb}{|V_{cb}|}
\newcommand{\vtd}{|V_{td}|}
\newcommand{\vub}{|V_{ub}|}
\newcommand{\vts}{|V_{ts}|}

\def\eps{\varepsilon}
\def\epe{\varepsilon'/\varepsilon}
\newcommand{\tev}{\, {\rm TeV}}
\newcommand{\gev}{\, {\rm GeV}}
\newcommand{\mev}{\, {\rm MeV}}

\newcommand{\RE}{{\rm Re}}

\newcommand{\be}{\begin{equation}}
\newcommand{\ee}{\end{equation}}
\newcommand{\bea}{\begin{eqnarray}}
\newcommand{\eea}{\end{eqnarray}}

\newcommand{\ba}{\begin{array}}
\newcommand{\ea}{\end{array}}

\newcommand{\ord}{{\cal O}}

\def\kpn{K^+\rightarrow\pi^+\nu\bar\nu}

\def\klpn{K_{L}\rightarrow\pi^0\nu\bar\nu}

\newcommand{\bsi}{B_6^{(1/2)}}
\newcommand{\bei}{B_8^{(3/2)}}
\newcommand{\kepe}{\kappa_{\varepsilon^\prime}}
\newcommand{\keps}{\kappa_{\varepsilon}}

\title{The Renaissance of Kaon Flavour Physics}

\ShortTitle{The Renaissance of Kaon Flavour Physics}

\author{\speaker{Andrzej J. Buras}%
 \thanks{FLAVOUR(267104)-ERC-122} \\
TUM-IAS, Lichtenbergstr. 2a, D-85748 Garching, Germany \\
Technical University Munich, Physics Department, D-85748 Garching, Germany,\\
E-mail: \email{aburas@ph.tum.de}}

\abstract{There is no doubt that in the coming years we will witness 
the renaissance of kaon flavour physics with crucial measurements of the branching ratios for the rare decays $\kpn$ and $\klpn$ that are very sensitive to new physics (NP) and are theoretically very clean. Simultaneously the role of $\epe$, $\varepsilon_K$, $\Delta M_K$, $K_L\to\mu^+\mu^-$ and $K_L\to\pi^0\ell^+\ell^-$in  searching for NP will significantly increase through their improved estimates within the SM. In fact the hints for NP contributing to $\epe$ 
 have been signalled  last year through improved estimates of hadronic 
matrix elements of QCD and electroweak penguin operators $Q_6$ and $Q_8$ 
by lattice QCD and large $N$ dual QCD approach. Also  recent 
increased tensions between $\varepsilon_K$ and $\Delta M_{s,d}$ mass differences in  $B_{s,d}^0-\bar B_{s,d}^0$ mixing within the SM and models with constrained 
MFV tell us that we should have tremendous fun in chasing NP through all 
these observables. The present talk discusses the hinted 
anomalies in $\epe$ and $\Delta F=2$ transitions and  summarizes possible 
 implications  of simultaneous anomalies in $\epe$ and $\varepsilon_K$ for $\kpn$, $\klpn$ and $\Delta M_K$ in models with tree-level $Z$ and $Z^\prime$ exchanges. The anomalies in $\epe$ and $\varepsilon_K$, if confirmed, would signal new 
sources of CP violation. This is in contrast to the recent anomalies in $B$ physics that signal  lepton flavour non-universalities and NP in  CP-conserving observables. Highlights from  recent flavour analyses of 331 models are briefly described. The correlations of kaon flavour observables with those in $B^\pm$, $B_{s,d}$ and $D$ meson systems will be crucial for the identification of flavour dynamics at very short distance scales.
 }

\FullConference{16th International Conference on B-Physics at Frontier 
                Machines\\
                 2-6 May 2016\\
                 Marseille, France}

\begin{document}

\section{Overture}
Flavour physics continues to play an important role 
in identifying new physics (NP) at short distance scales \cite{Buras:2013ooa,Isidori:2014rba,Buras:2015nta,Fleischer:2015mla}. The full picture will only 
be gained through the study of flavour violating processes in all meson 
systems and including also lepton flavour violation, breakdown of lepton flavour  universality, electric dipole moments and $(g-2)_{e,\mu}$. But in this talk I will discuss dominantly $K$ meson flavour physics which is gaining impetus 
through the recent developments mentioned in the abstract. 

In my talk at the EPS conference on HEP in Vienna last year I have discussed 
the following topics \cite{Buras:2015hna}:
\begin{itemize}
\item
The present status of $\kpn$ and $\klpn$ within the SM  \cite{Buras:2015qea} 
presenting parametric expressions for the branching ratios for these two 
decays in terms of the CKM input and the correlations between   $\kpn$ and 
$B_{s}\to\mu^+\mu^-$  and  between $\kpn$ and $\varepsilon_K$ in the SM. This was 
followed by the  results for both decays obtained in simplified models with flavour violating couplings of the SM $Z$ and of a heavy $Z^\prime$ \cite{Buras:2015yca}.
\item
The 2015 picture of quark flavour observables in the LHT model with T-parity after 
LHC Run 1 \cite{Blanke:2015wba}.
\item
New results on $\epe$ from lattice QCD  \cite{Blum:2015ywa,Bai:2015nea} 
and large $N$ approach \cite{Buras:2015xba} which have shown 
the emerging anomaly in $\epe$  \cite{Buras:2015yba} with its value in the SM being significantly below the experimental world average from NA48 \cite{Batley:2002gn} and KTeV \cite{AlaviHarati:2002ye,Abouzaid:2010ny} collaborations.
\item
The impact of these results on the correlation between  $\epe$ and $\klpn$ and $\kpn$ in the LHT model \cite{Blanke:2015wba} and the first look at such a correlation within 
simplified models in \cite{Buras:2015yca}.
\end{itemize}

The present talk can be considered as the continuation of the discussion in 
 \cite{Buras:2015hna} and will include the following topics
\begin{itemize}
\item
A significant tension between $\varepsilon_K$ and $\Delta M_{s,d}$ within the SM
 and models with constrained MFV (CMFV) \cite{Blanke:2016bhf} implied by 
 new lattice QCD results from Fermilab Lattice and MILC Collaborations \cite{Bazavov:2016nty}   on $B^0_{s,d}-\bar B^0_{s,d}$ hadronic matrix elements.
\item
A strategy for a systematic analysis of the implications of the $\epe$ anomaly, 
mentioned above, in conjunction with a possible anomaly in 
$\varepsilon_K$ on $\kpn$, $\klpn$ and $\Delta M_K$ in  models with flavour violating couplings of the SM $Z$ and of a heavy $Z^\prime$ \cite{Buras:2015jaq}.  
\item
New results on $\epe$ and their correlations with other flavour observables 
in 331 models based on the gauge group $SU(3)_C\times SU(3)_L\times U(1)_X$, 
recently presented in \cite{Buras:2015kwd,Buras:2016dxz}.
\end{itemize}
In view of space limitations I will try to reduce the overlap with \cite{Buras:2015hna} to a minimum so that potential readers are asked to look up also
the EPS15 writing \cite{Buras:2015hna}, in particular many references included there. For a very nice summary of other anomalies in flavour physics see
 \cite{Crivellin:2016ivz}.
\section{Tensions between $\varepsilon_K$ and $\Delta M_{s,d}$ in the SM and CMFV Models}\label{sec:3a}
The five observables of interest are in this case
\be\label{great5}
\Delta M_s,\quad \Delta M_d, \quad S_{\psi K_S},\quad S_{\psi \phi}, \quad \varepsilon_K
\ee
with $\Delta M_{s,d}$ being the mass differences in 
$B^0_{s,d}-\bar B^0_{s,d}$ mixings and $S_{\psi K_S}$ and $S_{\psi \phi}$ 
the corresponding mixing induced CP-asymmetries.
$\varepsilon_K$ describes the size of the indirect CP violation in $K^0-\bar K^0$ mixing. As seen in Table~\ref{tab:input} 
$\Delta M_{s,d}$ and  $\varepsilon_K$ 
are already known with impressive precision. The asymmetries  $S_{\psi K_S}$ and $S_{\psi \phi}$ are less precisely  measured but have the advantage of being subject to only very small hadronic uncertainties. The $K_L-K_S$ mass difference
$\Delta M_K$ is also very precisely determined but is subject to much larger theoretical uncertainties than the 
five observables in (\ref{great5}). Still as we will see in the next section 
it could play an important role in distinguishing between various BSM models 
in the future.

\begin{table}[!tb]
\center{\begin{tabular}{|l|l|}
\hline
 $\Delta M_s = 17.757(21) \,\text{ps}^{-1}$\hfill\cite{Amhis:2014hma}	&  $\Delta M_d = 0.5055(20) \,\text{ps}^{-1}$\hfill\cite{Amhis:2014hma}\\ 
$S_{\psi K_S}= 0.691(17)$\hfill\cite{Amhis:2014hma}
		&  $S_{\psi\phi}= 0.015(35)$\hfill \cite{Amhis:2014hma}\\
	$|V_{us}|=0.2253(8)$\hfill\cite{Agashe:2014kda} &
 $|\eps_K|= 2.228(11)\cdot 10^{-3}$\hfill\cite{Agashe:2014kda}\\
$F_{B_s}$ = $228.6(3.8)\mev$ \hfill \cite{Rosner:2015wva} & $F_{B_d}$ = $193.6(4.2)\mev$ \hfill \cite{Rosner:2015wva}  \\
$F_{B_s}\sqrt{\hat B_{B_s}}=(274.6\pm8.8)\mev$\hfill  \cite{Bazavov:2016nty} &
$F_{B_d} \sqrt{\hat B_{B_d}}=(227.7\pm 9.8)\mev$ \cite{Bazavov:2016nty}\\
$\eta_{cc}=1.87(76)$\hfill\cite{Brod:2011ty}  &  
 $\tilde\kappa_\varepsilon = 0.94(2)$\hfill \cite{Buras:2010pza}
\\	       
\hline
\end{tabular}  }
\caption {\textit{Values of the experimental and theoretical
    quantities used as input parameters.}}
\label{tab:input}
\end{table}

The hadronic uncertainties in   $\Delta M_{s,d}$ and $\varepsilon_K$  within 
the SM and CMFV models reside within a good approximation in the parameters
\be\label{hpar}
 F_{B_s}\sqrt{\hat B_{B_s}},\quad  F_{B_d} \sqrt{\hat B_{B_d}}, \quad \hat B_K, \quad \tilde\kappa_\varepsilon\,.
\ee
The first three represent the hadronic matrix elements of the relevant operators, while the last one long distance effects in $\varepsilon_K$.
During the last years these uncertainties decreased significantly.
Of particular interest is the recent improved precision on 
   $F_{B_s}\sqrt{\hat B_{B_s}}$ and $F_{B_d} \sqrt{\hat B_{B_d}}$,
 by the Fermilab Lattice and MILC Collaborations (Fermilab-MILC) \cite{Bazavov:2016nty} (see Table~\ref{tab:input}) as well as of their ratio
\be\label{xi}
\xi=\frac{F_{B_s}\sqrt{\hat B_{B_s}}}{F_{B_d}\sqrt{\hat B_{B_d}}}=1.206\pm0.019\,.
\ee
Its reduced uncertainty by a factor of three relative to previous results plays an  important  role in the analysis in \cite{Blanke:2016bhf}.
An extensive list of references to other
 lattice determinations of these parameters can be found in \cite{Bazavov:2016nty}. In particular the ETM Collaboration \cite{Carrasco:2013zta}  finds similar although less accurate results. Most impotantly
 their result for $\xi$  supports the one in (\ref{xi}).

Lattice QCD also made an impressive progress in
the determination of the parameter $\hat B_K$ which enters the evaluation of
$\varepsilon_K$. The most recent preliminary world 
average from FLAG reads $\hat B_K=0.7627(97)$ \cite{Vladikas:2015bra}, 
 very close to its large $N$  value $\hat B_K=0.75$ \cite{Gaiser:1980gx,Buras:1985yx}. In the dual QCD approach one finds $\hat B_K= 0.73\pm 0.02$ \cite{Buras:2014maa} and arguments can be provided that $\hat B_K$ cannot be larger than $0.75$ \cite{Gerard:2010jt,Buras:2014maa}.  Thus presently I quote
$\hat B_K=0.750 \pm 0.015$ in order to 
 encompass both large $N$ and lattice QCD values but  I expect that in the coming years  $\hat B_K$ from lattice QCD  will move below 0.75 as already signalled by the results in 
 \cite{Bae:2014sja,Carrasco:2015pra}.

The long distance contributions to $\varepsilon_K$,  represented by the departure of $\tilde\kappa_\varepsilon$ from unity, as given in Table~\ref{tab:input}, are reasonably well known so that
 at present the theoretical uncertainty in $\eps_K$  is dominated by the parameter $\eta_{cc} = 1.87 \pm 0.76$ \cite{Brod:2011ty} summarising NLO and NNLO QCD corrections to the charm quark contribution. From present perspective the only 
way this uncertainty can be reduced is through joint effort of lattice and 
perturbative QCD experts which would hopefully lead to an improved matching between Wilson coefficients and hadronic matrix elements in the charm contribution 
and to the improved estimate of long distance contributions. Such efforts would
also allow precise calculation of $\Delta M_K$ in the SM.

In order to demonstrate that the recent lattice results from \cite{Bazavov:2016nty} 
imply some tensions in the description of $\Delta F=2$ processes within the SM,
which are independent of some inconsistencies between inclusive and exclusive determinations of $\vub$ and $\vcb$, the  strategy of \cite{Blanke:2016bhf} is to leave the latter aside and use only results for $\Delta F=2$ observables.

\begin{figure}[!tb]
\centering
\includegraphics[width = 0.55\textwidth]{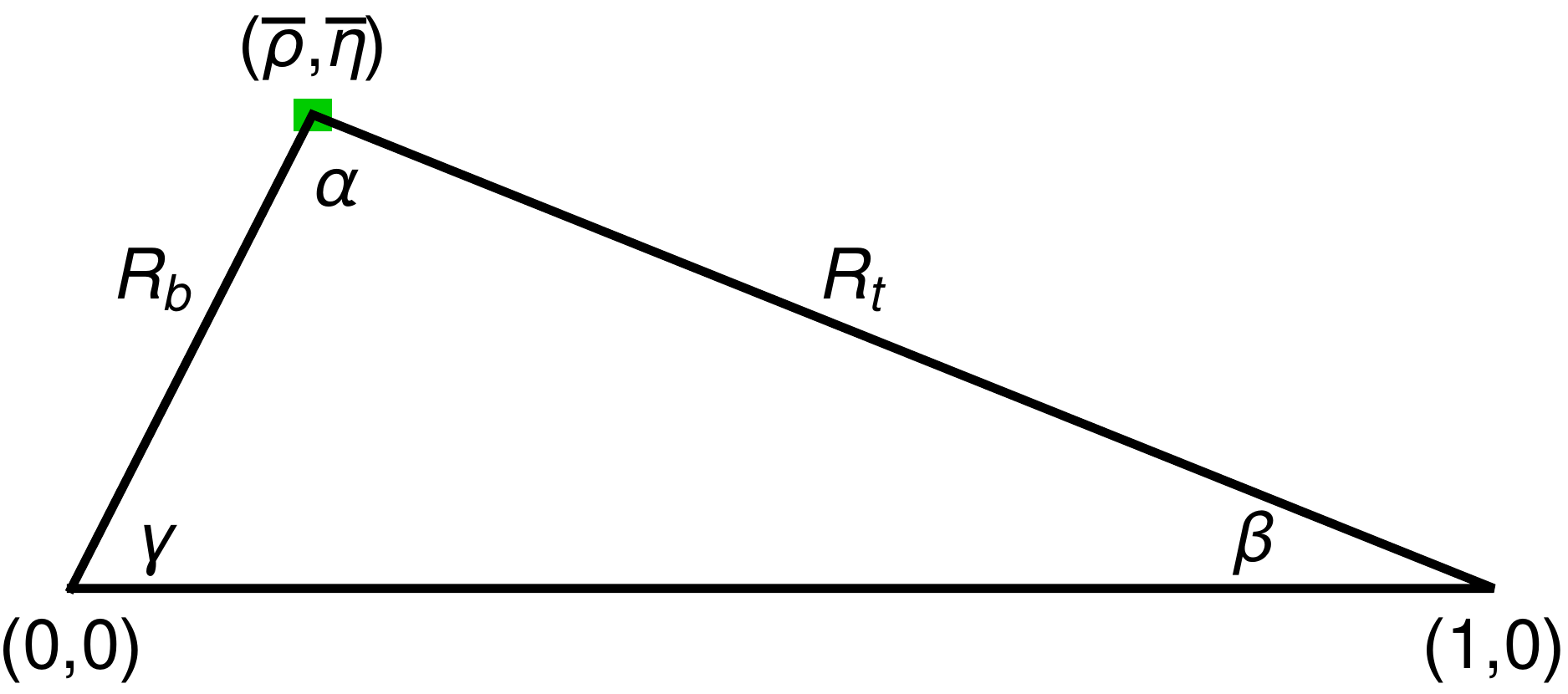}
 \caption{\it Universal Unitarity Triangle 2016. The green square at the apex 
of the UUT shows that the uncertainties in this triangle are very small. From 
\cite{Blanke:2016bhf}.}\label{UUTa}
\end{figure}

Now in the SM 
and more generally models with constrained minimal flavour violation (CMFV) \cite{Buras:2000dm,Buras:2003jf,Blanke:2006ig} it is possible to construct the so-called {\it universal unitarity triangle} (UUT) \cite{Buras:2000dm} 
 by using  only 
\be\label{UUT}
\frac{\Delta M_d}{\Delta M_s}, \qquad S_{\psi K_S}
\ee
and this in turn allows to determine $\vub/\vcb$ and the angle $\gamma$ in the 
UUT in  Fig.~\ref{UUTa}. As demonstrated 
in  \cite{Blanke:2016bhf} this determination is rather precise after 
the improved precision on $\xi$ in (\ref{xi})
\be\label{vubvcb}
\frac{\vub}{\vcb}= 0.0864\pm 0.0025 \qquad \gamma=(63.0\pm 2.1)^\circ\,.
\ee
We observe that the value of $\gamma$ is significantly below its central value 
of about $70^\circ$ from tree-level decays and has much smaller  uncertainty.

The important virtue of the determinations in (\ref{vubvcb}) and of the 
UUT  shown in Fig.~\ref{UUTa} is their universality within 
CMFV models. In the case of $\Delta F=2$ transitions in the 
down-quark sector various CMFV models can only be distinguished by the value of a
single flavour universal real one-loop  function, the 
box diagram function $S(v)$, with $v$ collectively denoting the parameters of a given 
CMFV model. This function enters universally $\varepsilon_K$, $\Delta M_s$ 
and $\Delta M_d$ and cancels out in the ratio in (\ref{UUT}). Therefore 
the resulting UUT is the same in all CMFV models. Its apex is determined
very precisely with $\bar\rho=0.170\pm0.013$ and $\bar\eta= 0.333\pm0.011$.
These values  differ significantly from those obtained in global fits \cite{Charles:2015gya,Bona:2006ah}, with the latter exhibiting smaller $\bar\rho$ and larger $\bar\eta$ values.

Moreover it can be shown that in these models $S(v)$ is bounded from below by its SM value \cite{Blanke:2006yh}
\be\label{BBBOUND}
S(v)\ge S_0(x_t)= 2.32 \,.
\ee

The tension between  $\varepsilon_K$ and  $\Delta M_s$ in the SM and CMFV 
can now be exhibited without the use of any  tree-level determinations, except for $\vus$,  by using the following two strategies  \cite{Blanke:2016bhf}:
\begin{itemize}
\item[{$S_1$:}]
{\boldmath $\Delta M_s$ \unboldmath} {\bf strategy}  in which the experimental value of 
$\Delta M_s$ is used to determine $\vcb$ as a function of $S(v)$, and $\varepsilon_K$ is then a derived quantity. 
\item[{$S_2$:}]
{\boldmath $\varepsilon_K$ \unboldmath}{\bf strategy} in which the experimental value of 
$\varepsilon_K$ is used, while $\Delta M_{s}$ is then a derived quantity and 
$\Delta M_d$ follows then from the determined UUT. 
\end{itemize}

Both strategies use the determination of the UUT {by means of (\ref{UUT})}
 and allow 
to determine the whole CKM matrix, in particular $\vts$, $\vtd$, $\vub$ and $\vcb$ as functions of $S(v)$. As seen in Table~\ref{tab:CKM}  their outcome is very different, 
which signals the tension between $\Delta M_{s,d}$ and $\varepsilon_K$ 
in this framework. Indeed
\begin{itemize}
\item
The lower bound in (\ref{BBBOUND}) implies in $S_1$ {\it upper bounds} on 
 $\vts$, $\vtd$, $\vub$ and $\vcb$ which are saturated in the SM, and 
in turn allows to derive an {\it upper bound} on $\varepsilon_K$ in CMFV models 
that is saturated in the SM but turns out to be significantly {\it below} the data
\be\label{ebound}
|\varepsilon_K|\le (1.64\pm0.25)\cdot 10^{-3}\,.
\ee
\item
The lower bound in (\ref{BBBOUND}) implies in $S_2$  also {\it upper bounds} on 
 $\vts$, $\vtd$, $\vub$ and $\vcb$ which are saturated in the SM. {However} 
the $S(v)$ dependence of these elements determined in this manner differs from
the one obtained in $S_1$, which in turn allows to derive  {\it lower bounds} on $\Delta M_{s,d}$ in CMFV models 
that are reached in the SM but turn out to be significantly {\it above} the data
\be\label{Msdbounds}
\Delta M_s \ge (21.1\pm 1.8)\text{ps}^{-1}, \qquad \Delta M_d \ge (0.600\pm0.064) \text{ps}^{-1}\, .
\ee
\end{itemize}

\begin{table}[!tb]
\centering
\begin{tabular}{|c|c|c|c|c|c|c|}
\hline
 $S_i$ & $\vts$ & $\vtd$ & $\vcb$ & $\vub$ &${\rm Im}\lambda_t$ & ${\rm Re}\lambda_t$        \\
\hline
$S_1$ &$39.0(13)$   & $8.00(29)$ & $39.7(13)$& $3.43(15)$ & $ 1.21(8)$& $-2.88(19)$\\
$S_2$ & $42.6(11)$  &$8.73(26) $ & $43.3(11) $& $3.74(14)$& $1.44(7)$ & $-3.42(18)$ \\
\hline
\end{tabular}
\caption{\it Upper bounds on CKM elements in units of $10^{-3}$ and of $\lambda_t=V_{ts}^*V_{td}$ in units of $10^{-4}$ obtained using strategies 
$S_1$ and $S_2$ as explained in the text. The bounds correspond to $S(v)=S_0(x_t)$. 
}\label{tab:CKM}
\end{table}

\begin{figure}[!tb]
\centering
\includegraphics[width = 0.49\textwidth]{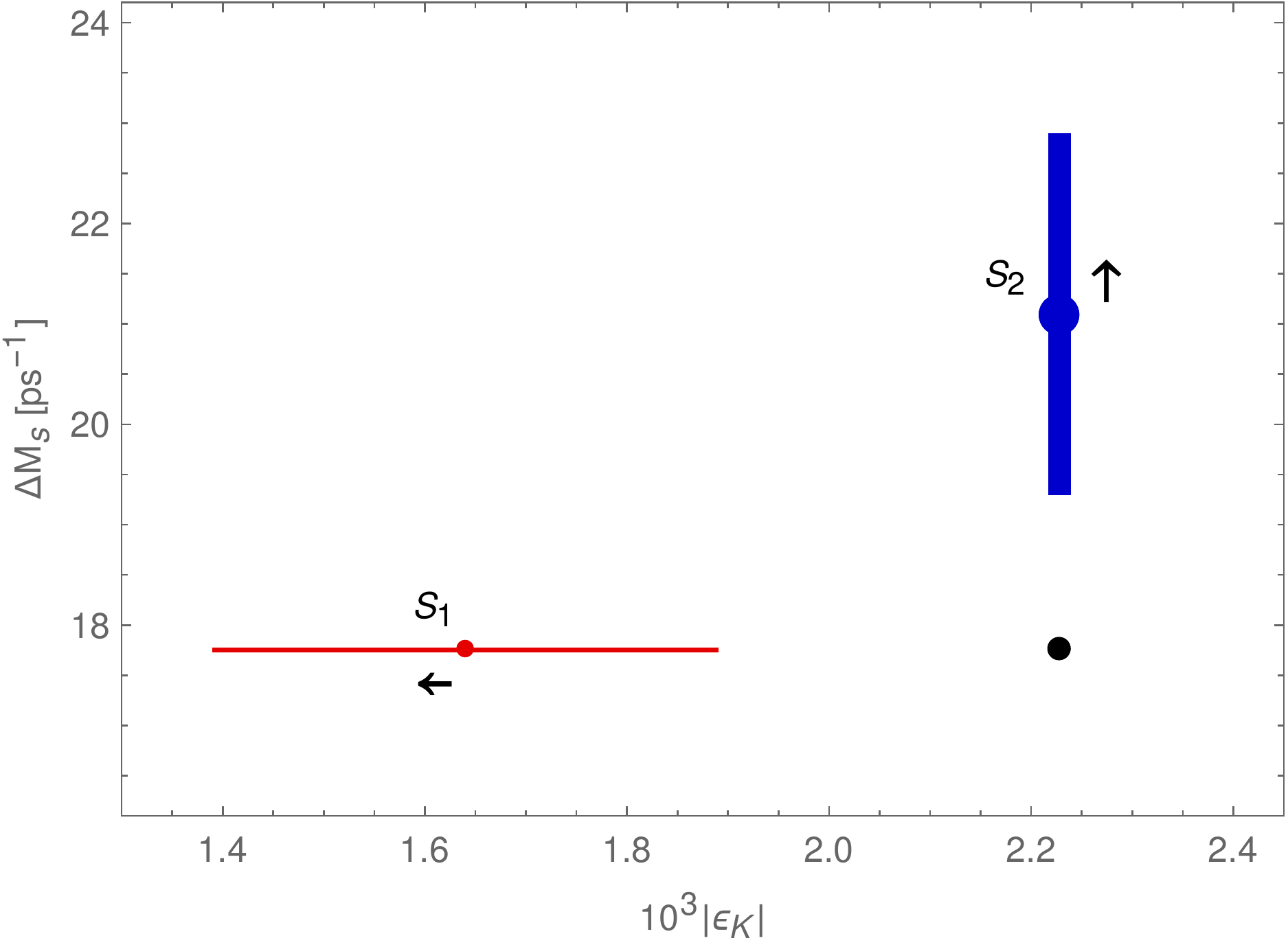}\hfill
\includegraphics[width = 0.49\textwidth]{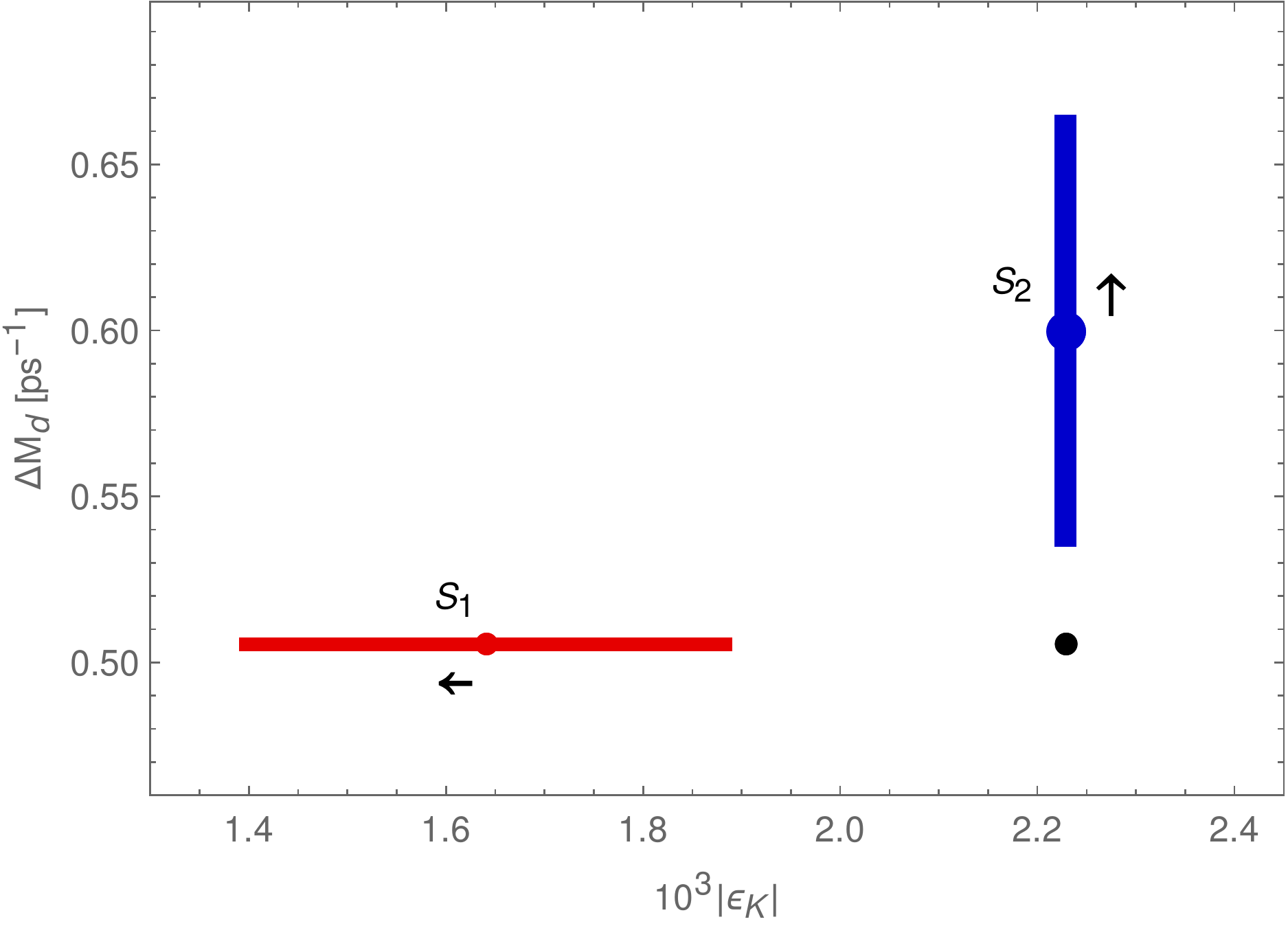}
 \caption{\it  $\Delta M_{s,d}$ and  $\varepsilon_K$ obtained from the strategies
$S_1$ and $S_2$ for $S(v)=S_0(x_t)$, at which the upper bound on $\varepsilon_K$ in $S_1$ and lower bound on  $\Delta M_{s,d}$ in $S_2$ are obtained.
The arrows show how the red and blue regions move with increasing $S(v)$. {The black dot represents the experimental values. From \cite{Blanke:2016bhf}.}}\label{eKvsDMsdplots}
\end{figure}

 The resulting low value of $\varepsilon_K$ in $S_1$
can be naturally raised in CMFV models by enhancing the value of $S(v)$ 
or/and increasing the value of $\vcb$. 
However, as pointed out in \cite{Buras:2011wi,Buras:2012ts,Buras:2013raa}, this spoils 
the agreement with the data on $\Delta M_{s,d}$, signalling
 the  tension between $\Delta M_{s,d}$ and $\varepsilon_K$ in CMFV models. 
We conclude therefore, as already indicated by the analysis in  \cite{Buras:2013raa}, that it is impossible within CMFV 
models to obtain a simultaneous satisfactory agreement of $\Delta M_{s,d}$ and $\varepsilon_K$ with the data.
In the context of the strategies $S_1$ and $S_2$, the tension 
between $\Delta M_{d,s}$ and $\varepsilon_K$ is summarized by the plots
of $\Delta M_{s,d}$ vs. $\varepsilon_K$ in Fig.~\ref{eKvsDMsdplots}.

This analysis will be generalized in the future
through the 
measurements of the branching ratios $\kpn$ and $\klpn$. 
As pointed out already in \cite{Buras:1994ec} precise measurements of both $K\to\pi\nu\bar\nu$ branching ratios would offer the determination 
of the unitarity triangle which could be compared with the one extracted these
days dominantly from $B$ physics. In particular as demonstrated in 
 \cite{Buchalla:1994tr,Buras:2001af} within the SM and models with MFV, 
rather precise determination of $\sin 2\beta$ without the usual QCD 
penguin pollution and almost independently of $\vcb$ can be obtained. Analytic 
expressions for the parameters $\bar\varrho$ and $\bar\eta$ in terms of 
the branching ratios for $\kpn$ and $\klpn$ can be found in these papers and 
numerical analyses have been presented by us in several papers since then.
Recently also the authors of \cite{Lehner:2015jga} have shown that the inclusion of $\varepsilon_K$ and $\epe$ in addition to  $\kpn$ and $\klpn$ could help 
to determine such K-triangle.

\section{$\epe$, $\varepsilon_K$, $\kpn$, $\klpn$ and $\Delta M_K$}\label{sec:4}
\subsection{$\epe$ Striking Back}
The most recent estimate of $\epe$ in the SM reads
\cite{Buras:2015yba} 
\be\label{LBGJJ}
   \epe = (1.9 \pm 4.5) \times 10^{-4} \,,
\ee
roughly $3\sigma$ away from 
the world average from NA48 \cite{Batley:2002gn} and KTeV
\cite{AlaviHarati:2002ye,Abouzaid:2010ny} collaborations 
\be\label{EXP}
(\epe)_\text{exp}=(16.6\pm 2.3)\times 10^{-4} \,.
\ee
It is based on the RBC-UKQCD collaboration results on the relevant hadronic matrix elements of the QCD penguin operator $Q_6$ \cite{Bai:2015nea}
and electroweak penguin operator $Q_8$ \cite{Blum:2015ywa}. 
These results imply the following values 
for the known parameters $\bsi$ and $\bei$ \cite{Buras:2015yba,Buras:2015qea}
\be\label{Lbsi}
\bsi=0.57\pm 0.19\,, \qquad \bei= 0.76\pm 0.05\,, \qquad (\mbox{RBC-UKQCD}),
\ee
significantly below their values in the strict large $N$ limit 
\cite{Buras:1985yx,Bardeen:1986vp,Buras:1987wc} 
\be\label{LN}
\bsi=\bei=1, \qquad {\rm (large~N~Limit)}\,.
\ee

This suppression of $\bsi$ and $\bei$ below the unity can be understood in the 
the dual QCD approach  \cite{Buras:2015xba} as the effect of the meson 
evolution from scales  $\mu=\ord(m_\pi,m_K)$  at which (\ref{LN}) is valid to 
 $\mu=\ord(1\gev)$ at which Wilson coefficients of $Q_6$ and $Q_8$ are 
evaluated. This evolution has to be matched to the perturbative quark evolution
for scales higher than $1\gev$ and in fact the supressions in question
and the property that $\bsi$ is more strongly suppressed than $\bei$ are 
consistent with the perturbative evolution of these parameters above  
$\mu=\ord(1\gev)$. Thus we are rather confident that \cite{Buras:2015xba}
\be\label{NBOUND}
\bsi\le \bei < 1 \, \qquad ({\rm dual~QCD}).
\ee
Explicit calculation in this approach gives $B_8^{(3/2)}(m_c)=0.80\pm 0.10$.
The result for $\bsi$ is less precise but  in agreement with
(\ref{Lbsi}). For further details, see \cite{Buras:2015xba}.

In this context it should be emphasized that in the past values $\bsi=\bei=1.0$ 
have been combined in phenomenological applications with the Wilson coefficients evaluated at scales $\mu=\ord(1\gev)$. The discussion above shows that this 
is incorrect. The meson evolution from  $\mu=\ord(m_\pi,m_K)$  to $\mu=\ord(1\gev)$ 
has to be performed and this effect turns out to be stronger than the 
scale dependence of $\bsi$ and $\bei$ in the perturbative regime, where the 
scale dependence of these parameters is very weak.

Additional support for the small value of $\epe$ in the SM comes from the recent reconsideration of the role of final state interactions (FSI) in $\epe$ 
\cite{Buras:2016fys}. Already long time ago the chiral perturbation theory practitioners put forward the idea
that both the amplitude ${\rm Re}A_0$, governed by the current-current operator  $Q_2-Q_1$ and the $Q_6$ contribution to the ratio $\epe$ could be 
enhanced significantly through FSI in a correlated manner 
\cite{Antonelli:1995gw,Bertolini:1995tp,Pallante:1999qf,Pallante:2000hk,Buchler:2001np,Buchler:2001nm,Pallante:2001he} bringing the values of $\epe$ close to its experimental value.
However, as shown recently in \cite{Buras:2016fys}
 FSI are likely to be important for the $\Delta I=1/2$  rule, in agreement with \cite{Antonelli:1995gw,Bertolini:1995tp,Pallante:1999qf,Pallante:2000hk,Buras:2000kx,Buchler:2001np,Buchler:2001nm,Pallante:2001he}, but much less relevant  for $\epe$. It should also be
emphasized that the latter authors did not include the meson evolution of $\bsi$ and $\bei$ in their
analysis and already this effect would significantly lower their predictions for  $\epe$.

It should finally be noted that 
even without  lattice results, varying all input parameters, the bound 
in (\ref{NBOUND}) implies the upper bound on $\epe$ in the SM
 \be\label{BoundBGJJ}
(\epe)_\text{SM}<(8.6\pm 3.2) \times 10^{-4} \,,
\ee
still $2\,\sigma$  below the experimental data. On the other hand employing the lattice value for $\bei$ in (\ref{Lbsi}) and
$\bsi=\bei=0.76$, one obtains  $(6.0\pm 2.4)\times 10^{-4}$  instead of (\ref{BoundBGJJ}). We observe then that even for these values of $\bsi$ and $\bei$ the SM predictions
for $\epe$ are significantly below the data. 

These results give strong motivation for searching for NP which could enhance 
$\epe$ above its SM value. On the other hand in the case of
 $\varepsilon_K$ 
one can only claim, as discussed in the previous section, that there is a tension between $\varepsilon_K$ and $\Delta M_{s,d}$ in the whole class of
 CMFV models. Moreover, there is some tendency that $\varepsilon_K$ in 
the SM is below the data 
\cite{Lunghi:2008aa,Buras:2008nn,Bona:2006ah,Charles:2015gya,Bailey:2015frw},
but certainly one cannot talk presently about
an anomaly in $\varepsilon_K$. 

Still it is interesting to ask next what would 
be the implications of combined anomalies in $\epe$ and $\varepsilon_K$ for 
rare decays $\kpn$ and $\klpn$. This question can only be answered in 
a concrete NP scenario. Here we will summarize such implications 
 in a number of simple scenarios with FCNCs appearing already at tree-level 
and being mediated by $Z$ boson exchanges or  $Z^\prime$ exchanges. 

In this context it should be recalled that one of the reasons for a large uncertainty in the SM prediction for $\epe$ is the strong cancellation between QCDP and EWP contributions.
As stressed in  \cite{Buras:2015jaq}, beyond the SM, quite generally either 
EWP or QCDP dominate NP contributions to $\epe$ and theoretical uncertainties 
in these contributions are 
much smaller because no cancellations take place. We refer to  \cite{Buras:2015jaq} for the discussion of this point.

\subsection{Strategy}\label{sec:strategy}
In order to investigate the implications of anomalies in $\epe$ and $\varepsilon_K$ for rare decays $\kpn$ and $\klpn$ in a systematic fashion a strategy,
consisting of four steps, has been proposed in \cite{Buras:2015jaq}. We 
will describe now these steps. Subsequently we will collect the lessons
from this study. 

In this strategy the central role is played by $\epe$ and 
$\varepsilon_K$ for which in the presence of NP contributions we have
\be\label{GENERAL}
\frac{\varepsilon'}{\varepsilon}=\left(\frac{\varepsilon'}{\varepsilon}\right)^{\rm SM}+\left(\frac{\varepsilon'}{\varepsilon}\right)^{\rm NP}\,, \qquad \varepsilon_K\equiv 
e^{i\varphi_\eps}\, 
\left[\varepsilon_K^{\rm SM}+\varepsilon^{\rm NP}_K\right] \,.
\ee
Therefore, it should be emphasized at this point that the anomalies in both quantities, 
if confirmed, would signal the presence of new sources of CP-violation beyond 
the CKM framework. This should be contrasted with the recent anomalies 
in $B\to D(D^*)\tau\nu_\tau$ and $B\to K(K^*)\ell^+\ell^-$ that signal 
lepton flavour non-universalities and new sources of flavour violation in 
CP-conserving observables.

As in the case of $\epe$ and $\varepsilon_K$, in contrast to most $B$-physics 
observables, the intereference between SM and NP contributions is totally negligible, one can
fully concentrate the discussion on NP contributions. Therefore in order to identify the pattern of NP contributions to flavour observables
implied by the $\epe$ anomaly in a transparent manner, we can  proceed 
in a given NP model as follows:

{\bf Step 1:} We assume that NP provides a positive shift in $\epe$: 
\be\label{deltaeps}
\left(\frac{\varepsilon'}{\varepsilon}\right)^{\rm NP}= \kepe\cdot 10^{-3}, \qquad   0.5\le \kepe \le 1.5,
\ee
with the range for $\kepe$ indicating the required size of this contribution. 
This step 
determines the imaginary parts of flavour-violating $Z$ or $Z^\prime$ couplings to quarks as functions of $\kepe$.

{\bf Step 2:} Knowing that NP contribution to $\varepsilon_K$ is proportional to  the product of the real and imaginary parts of the couplings in question 
 we assume that in addition to the $\epe$ 
anomaly, NP can also affect the parameter  $\varepsilon_K$. This then 
allows to determine the relevant real parts of the couplings 
involved, in the presence of the imaginary part determined from $\epe$. 
We describe 
this effect by the parameter $\keps$  so that now in addition 
to (\ref{deltaeps}) we study the implications of the shift in $\varepsilon_K$ due to NP
\be \label{DES}
(\varepsilon_K)^{\rm NP}= \keps\cdot 10^{-3},\qquad 0.1\le \keps \le 0.4 \,.
\ee
The positive sign of $\keps$  is motivated by the discussions in the previous 
section but we cannot exclude that $\keps$ is smaller and even slightly negative. But one should note that in the case of $\keps=0$ but $\kepe\not=0$, the 
flavour-violating $Z$ and $Z^\prime$ couplings must be imaginary implying 
a strong correlation between the branching ratios for $\kpn$ and $\klpn$ and a
{\it negative} shift in $\Delta M_K$ from NP \cite{Buras:2015jaq}.

{\bf Step 3:} In view of the uncertainty in $\kepe$ one can first set several
input  parameters  to their central values. In particular 
for the SM contributions to rare decays  the CKM factors have been set 
in \cite{Buras:2015jaq} to 
\be\label{CKMfixed}
{\rm Re}\lambda_t=-3.0\cdot 10^{-4}, \qquad {\rm Im}\lambda_t=1.4\cdot 10^{-4}\,.
\ee
 These values are in the ballpark of present  estimates obtained by UTfit \cite{Bona:2006ah} and CKMfitter \cite{Charles:2015gya} collaborations. For this 
choice of CKM parameters the central value of the resulting $\varepsilon_K^{\rm SM}$ is $1.96\cdot 10^{-3}$. With the experimental value of $\varepsilon_K$  
 this  implies $\keps=0.26$.  But it is useful to vary 
$\keps$ while keeping the values in (\ref{CKMfixed}) fixed as NP contributions 
do not depend on them but are sensitive functions of $\keps$.

{\bf Step 4:} Having fixed the flavour violating couplings of $Z$ or $Z^\prime$ 
in this manner, one can express NP contributions to the branching ratios for $\kpn$, $\klpn$ 
and $K_L\to\mu^+\mu⁻$ and to $\Delta M_K$ in terms of $\kepe$ and $\keps$. 
Explicit formulae can be found in \cite{Buras:2015jaq}. In this manner one 
can directly study the impact of $\epe$ and $\varepsilon_K$ anomalies in $Z$ and $Z^\prime$ scenarios on these four observables.  The pattern of flavour violation  depends in a given NP scenario on the relative size of real and imaginary parts of the couplings 
involved and we will see this explicitly in the lessons below.

The present strategy above assumes that the progress in the evaluation of $\epe$
 in the SM will be faster than experimental information on $\kpn$. If in 2018 
the situation will be reverse, it will be better to choose as variables 
$\keps$ and $R^{\nu\bar\nu}_+$ defined by
\be\label{Rnn+}
R^{\nu\bar\nu}_+\equiv\frac{\mathcal{B}(\kpn)}{\mathcal{B}(\kpn)_\text{SM}},\qquad
R^{\nu\bar\nu}_0\equiv \frac{\mathcal{B}(\klpn)}{\mathcal{B}(\klpn)_\text{SM}}\,
\ee
with $R^{\nu\bar\nu}_0$ to be determined in the next decade.
Presently one can only 
provide $R^{\nu\bar\nu}_+$ as a function of $\kepe$ for fixed values of $\keps$ using the  strategy above. But knowing  $R^{\nu\bar\nu}_+$  better than $\epe$ in the SM will allow us to read off from the  plots presented in  \cite{Buras:2015jaq} and 
here
 the favourite range for $\kepe$ in a given NP scenario for a given $\keps$ and 
the diagonal lepton or quark couplings of $Z^\prime$. The dictionary 
for doing this can be found in  \cite{Buras:2015jaq}.
 Note that knowing $R^{\nu\bar\nu}_+$ will allow us to obtain 
$R^{\nu\bar\nu}_0$ directly from the plots in \cite{Buras:2015jaq}, using the 
value of $\kepe$ extracted from  $R^{\nu\bar\nu}_+$ and $\keps$. 
Clearly, when $R^{\nu\bar\nu}_0$ will also be known the analysis will be rather
constrained. 

\subsection{Lessons on  NP  Patterns in $Z$ Scenarios}\label{LessonsZ}

We will now summarize the lessons in $Z$ and $Z^\prime$ scenarios obtained in
\cite{Buras:2015jaq}. To this end we define flavour violating couplings 
$\Delta^{sd}_{L,R}(Z)$ by \cite{Buras:2012jb}
\be\label{Zcouplings}
i\mathcal{L}(Z)=i\left[\Delta_L^{sd}(Z)(\bar s\gamma^\mu P_Ld)+\Delta_R^{sd}(Z)(\bar s\gamma^\mu P_R d)\right] Z_\mu,\qquad P_{L,R}=\frac{1}{2}(1\mp \gamma_5)\,
\ee
with analogous definitions for $Z^\prime$ couplings.

The flavour violating couplings of $Z$ in a given model can be left-handed (LH), 
right-handed (RH) or can be an arbitrary linear combination of LH
and RH couplings, to be termed general scenario.
In \cite{Buras:2015jaq} all these possibilities have been considered. In 
what follows we will use for the first scenarios the abbreviations:
\be
{\rm LHS\, \equiv\, left-handed~scenario},\qquad {\rm RHS\, \equiv\, right-handed~scenario}.
\ee

The summary of the lessons is rather brief. On the other hand the 
presentation in \cite{Buras:2015jaq} is very detailed with numerous analytic
expressions.

{\bf Lesson 1:}
In the LHS, a given request for the enhancement of $\epe$ determines 
the coupling ${\rm Im} \Delta_{L}^{s d}(Z)$.

{\bf Lesson 2:}
In LHS there is a direct unique implication of an enhanced $\epe$ on $\klpn$: {\it suppression} of
  $\mathcal{B}(\klpn)$. This property is known from NP scenarios in which 
 NP to $\klpn$ and $\epe$ enters dominantly through the modification of 
$Z$-penguins. The known flavour diagonal $Z$ couplings to quarks and leptons 
and the sign of the matrix element $\langle Q_8\rangle_2$ determines this anticorrelation
which has been verified in all models with only LH flavour-violating $Z$ couplings. Fig.~1 in \cite{Buras:2015jaq} shows this anticorrelation.

\begin{figure}[!tb]
 \centering
\includegraphics[width = 0.45\textwidth]{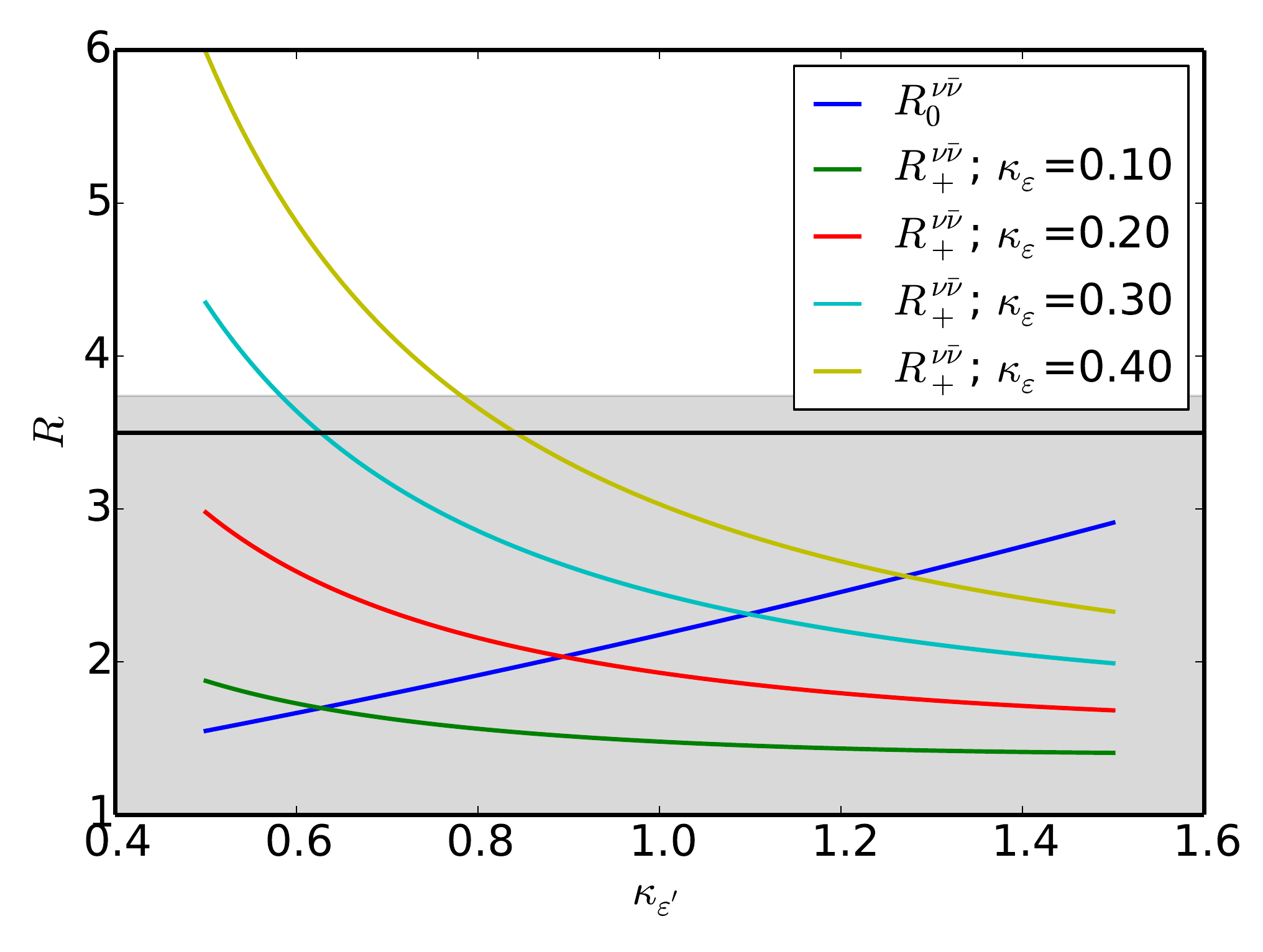}
\includegraphics[width = 0.45\textwidth]{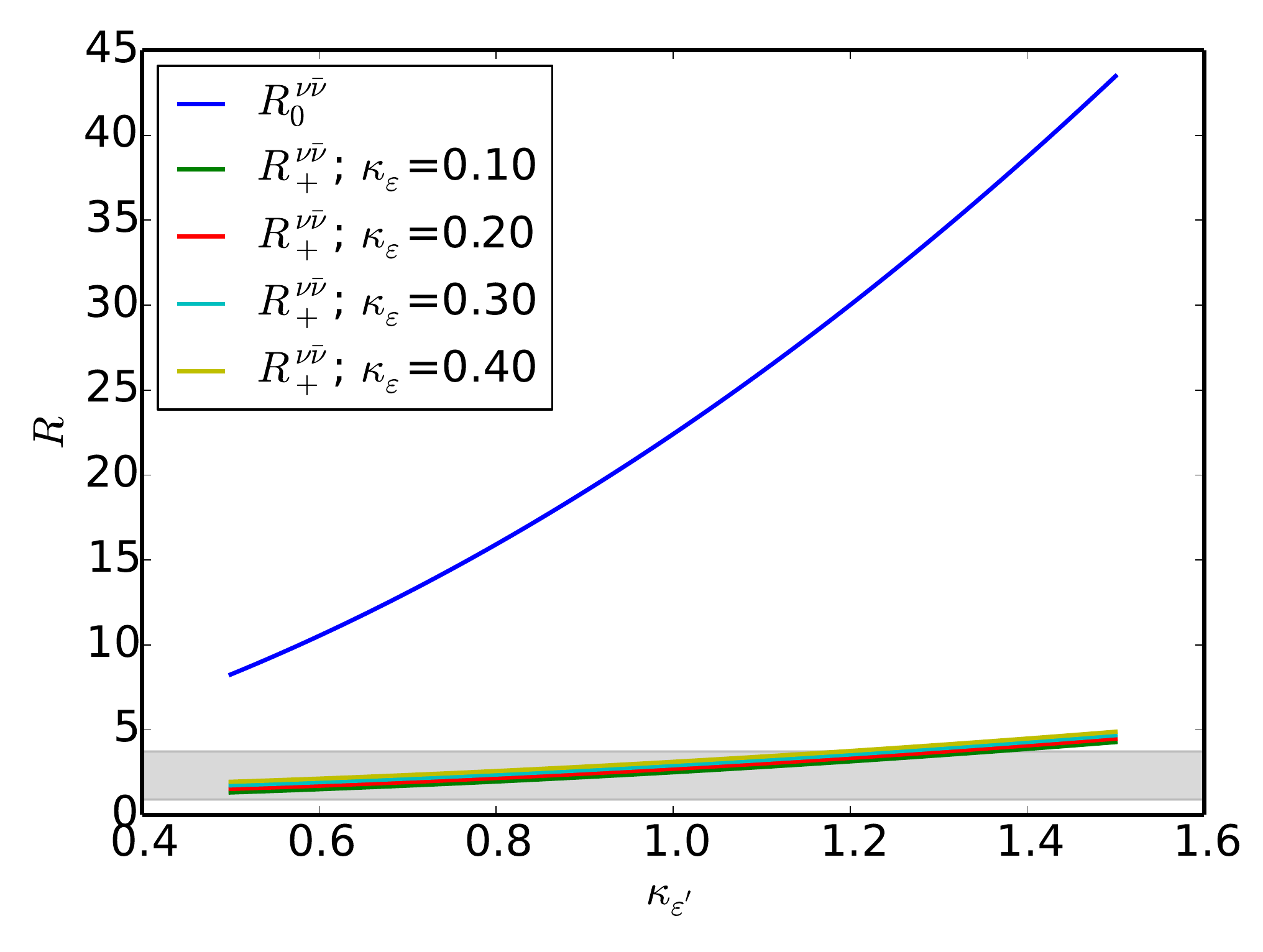}
\caption{ \it $R_0^{\nu\bar\nu}$ and  $R_+^{\nu\bar\nu}$, as functions 
of $\kepe$ for $\keps=0.1,\,0.2,\, 0.3,\, 0.4$  in the case of the dominance of real parts of the flavour-violating $Z$ couplings (example 1, left panel) and 
for the case of the dominance of imaginary parts of the flavour-violating $Z$ couplings (example 2, right panel). The horizontal 
{\it black} line in the left panel corresponds to the upper bound from $K_L\to\mu^+\mu^-$. The experimental $1\sigma$ range for  $R_+^{\nu\bar\nu}$ is displayed by the grey band. From \cite{Buras:2015jaq}.
}\label{R2a}~\\[-2mm]\hrule
\end{figure}

\begin{figure}[!tb]
 \centering
\includegraphics[width = 0.45\textwidth]{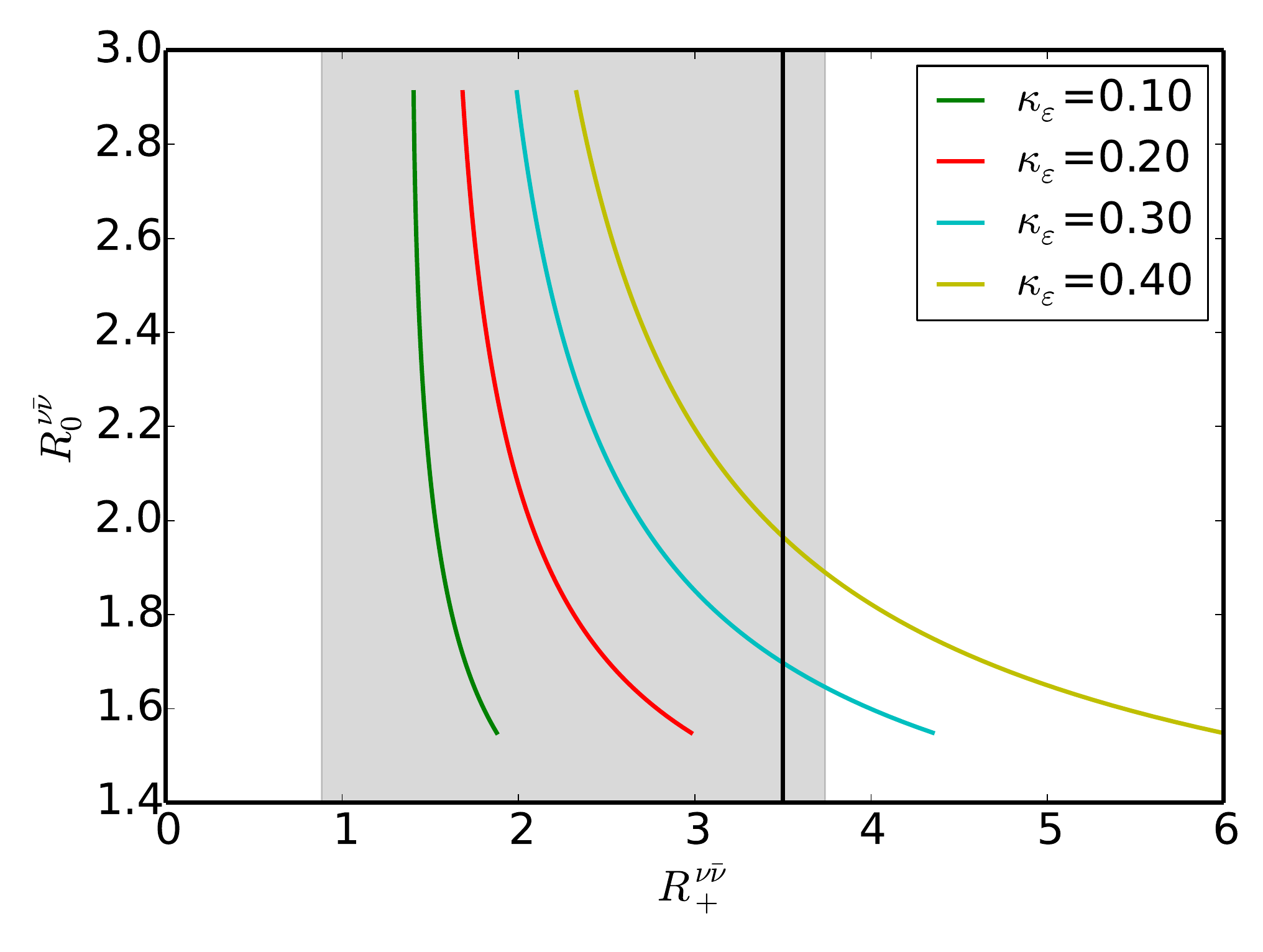}
\includegraphics[width = 0.45\textwidth]{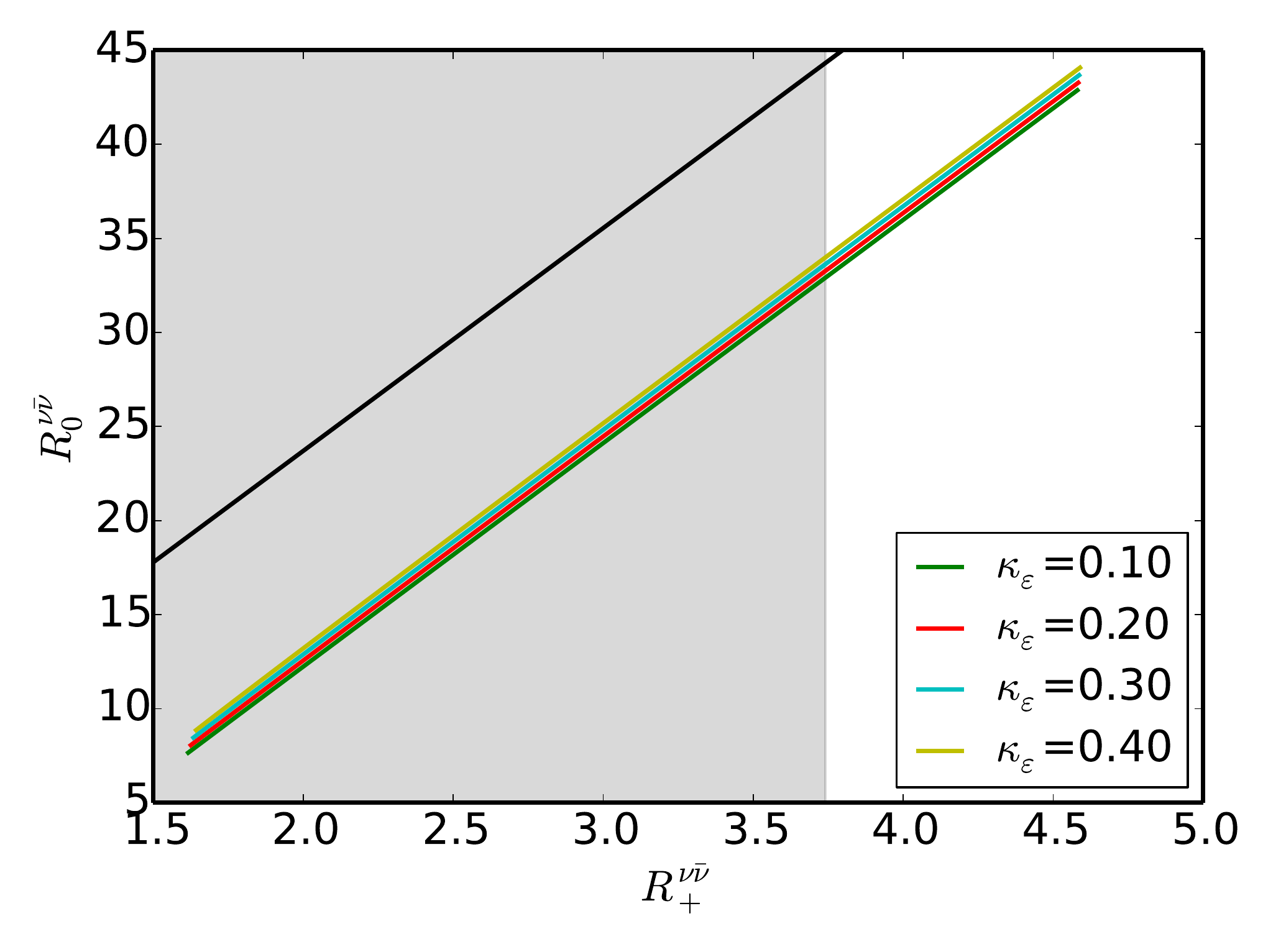}
\caption{ \it $R_0^{\nu\bar\nu}$ vs  $R_+^{\nu\bar\nu}$  
for $\keps=0.1,\,0.2,\, 0.3,\, 0.4$ for the example 1 (left panel) and the
example 2 (right panel) varying $0.5\le\kepe\le 1.5$. The vertical 
{\it black} line in the left panel corresponds to the upper bound from $K_L\to\mu^+\mu^-$. The dependence on $\keps$ in the right panel is negligible and the {\it black} line 
represents the Grossman-Nir (GN) bound \cite{Grossman:1997sk}. The experimental $1\sigma$ range for  $R_+^{\nu\bar\nu}$ is displayed by the grey band.  From \cite{Buras:2015jaq}.
}\label{R4ab}~\\[-2mm]\hrule
\end{figure}

{\bf Lesson 3:} 
The imposition of the $K_L\to\mu^+\mu^-$ constraint in LHS 
determines the range for
${\rm Re} \Delta_{L}^{s d}(Z)$ which with the already fixed ${\rm Im} \Delta_{L}^{s d}(Z)$ allows to calculate the shifts in $\varepsilon_K$ and $\Delta M_K$. 
These shifts turn out to be very small for $\varepsilon_K$ and negligible for 
 $\Delta M_K$. Therefore unless loop contributions from physics generating 
$ \Delta_{L}^{s d}(Z)$ play a significant role in both quantities, the SM predictions
 for  $\varepsilon_K$ and $\Delta M_K$ must agree well with data for this NP scenario to survive. In  models with vector-like quarks (VLQs) box diagrams with VLQs 
can indeed provide contributions to  $\varepsilon_K$ and $\Delta M_K$ that 
are larger than coming from tree-level $Z$-exchange when the masses
of VLQs are far above $1~\tev$ \cite{Ishiwata:2015cga}.

{\bf Lesson 4:}
With fixed ${\rm Im} \Delta_{L}^{s d}(Z)$ and the allowed range 
for ${\rm Re} \Delta_{L}^{s d}(Z)$, the range for  $\mathcal{B}(\kpn)$ 
can be obtained. But in view of uncertainties in the  $K_L\to\mu^+\mu^-$ constraint \cite{Isidori:2003ts} both an enhancement and a suppression of  $\mathcal{B}(\kpn)$ are possible 
and no specific pattern of correlation between  $\mathcal{B}(\klpn)$ and 
 $\mathcal{B}(\kpn)$ is found. In the absence of a relevant $\varepsilon_K$ 
constraint this is consistent with the general analysis in \cite{Blanke:2009pq}.
$\mathcal{B}(\kpn)$  can be enhanced by a factor of $2$ at most.

{\bf Lesson 5:}
Analogous pattern is found in RHS, although the numerics is different. See
  Fig.~1 in \cite{Buras:2015jaq}. First  due the increased value of the Wilson coefficients of the relevant $Q_8^\prime$ operator relative to the one of $Q_8$ operator,
the suppression of $\mathcal{B}(\klpn)$ for a given $\kepe$ is smaller.
Moreover,  the flip of the sign in NP contribution to $K_L\to \mu^+\mu^-$  allows 
for larger enhancement of  $\mathcal{B}(\kpn)$, a property first pointed 
out in \cite{Blanke:2008yr}. An enhancement of $\mathcal{B}(\kpn)$  up to a factor of  $5.7$ 
is possible.

{\bf Lesson 6:}
In a general $Z$ scenario with LH and RH flavour-violating couplings the pattern of NP effects 
changes because of the appearance of 
LR operators dominating NP contributions to  $\varepsilon_K$ and $\Delta M_K$. 
Consequently for a large range of parameters  
these two quantities, in particular  $\varepsilon_K$, provide stronger 
constraint on ${\rm Re} \Delta_{L,R}^{s d}(Z)$ than $K_L\to\mu^+\mu^-$. But 
the main virtue of the general scenario is the possibility of enhancing 
simultaneously $\epe$, $\varepsilon_K$, $\mathcal{B}(\kpn)$ and $\mathcal{B}(\klpn)$ which is not possible in LHS and RHS. Thus the presence of both 
LH and RH flavour-violating currents is essential for obtaining
simultaneously the enhancements in question. The correlations 
between $\epe$ and $\kpn$ and $\klpn$ depend sensitively on the ratio of 
real and imaginary parts of the flavour-violating couplings involved. In the left panel of Fig.~\ref{R2a} we show 
$R_0^{\nu\bar\nu}$ and  $R_+^{\nu\bar\nu}$, as functions 
of $\kepe$ for the case of the dominance of real parts (example 1) and in the 
right panel for the case of the dominace of imaginary parts (example 2).
In Fig.~\ref{R4ab} we show the correlations between $R_0^{\nu\bar\nu}$ and  $R_+^{\nu\bar\nu}$ for these two examples.
We observe that in the example 2 the correlation 
takes place along the line parallel to the line representing Grossman-Nir bound  \cite{Grossman:1997sk}. Otherwise, as is the case of example 1, the correlation is rather different. 

But the main message from this analysis is that in the presence of both LH
and RH flavour-violating couplings of $Z$ to quarks, large departures from 
SM predictions for $\kpn$ and $\klpn$ are possible.

\begin{figure}[!tb]
 \centering
\includegraphics[width = 0.45\textwidth]{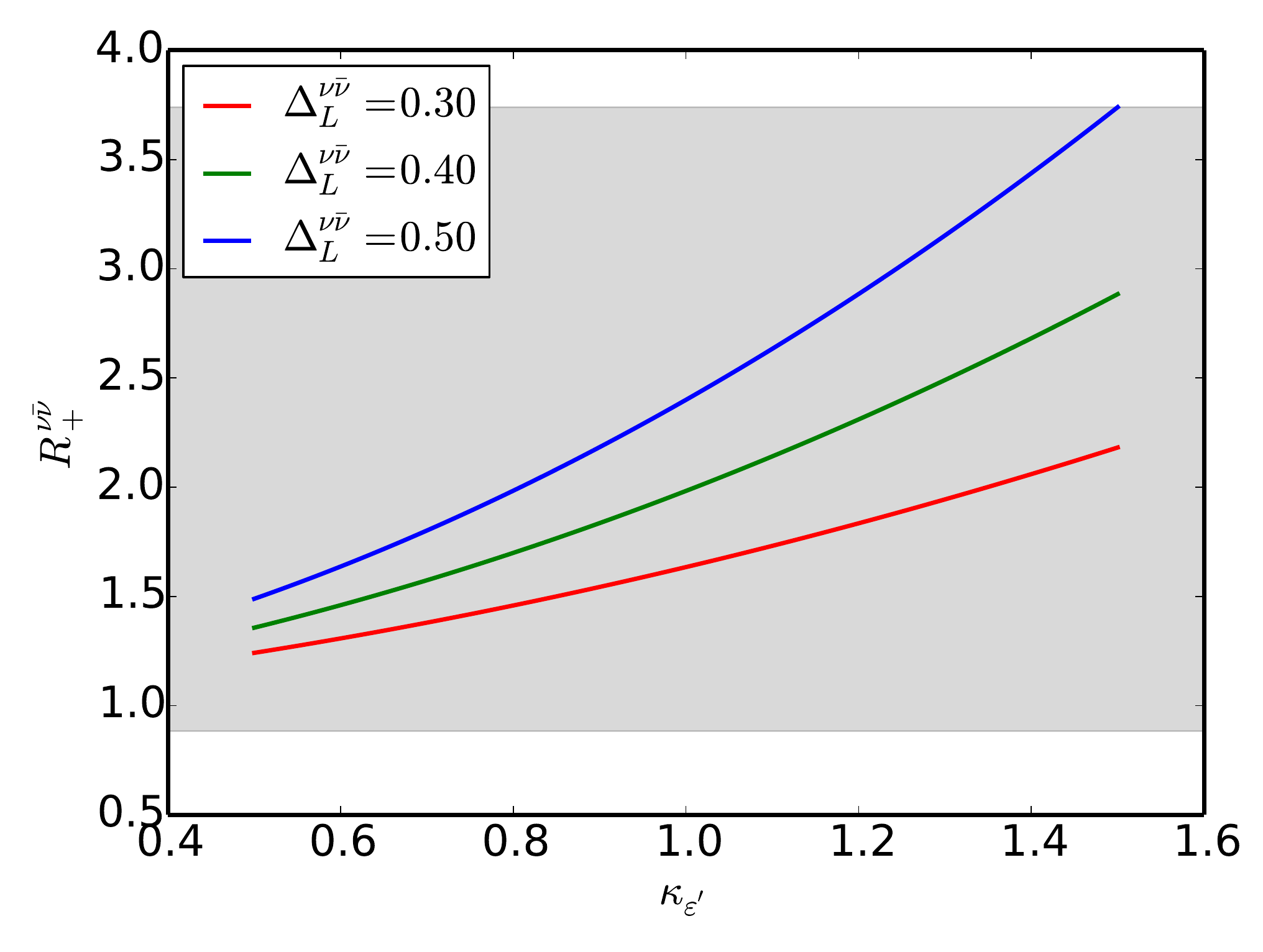}
\includegraphics[width = 0.45\textwidth]{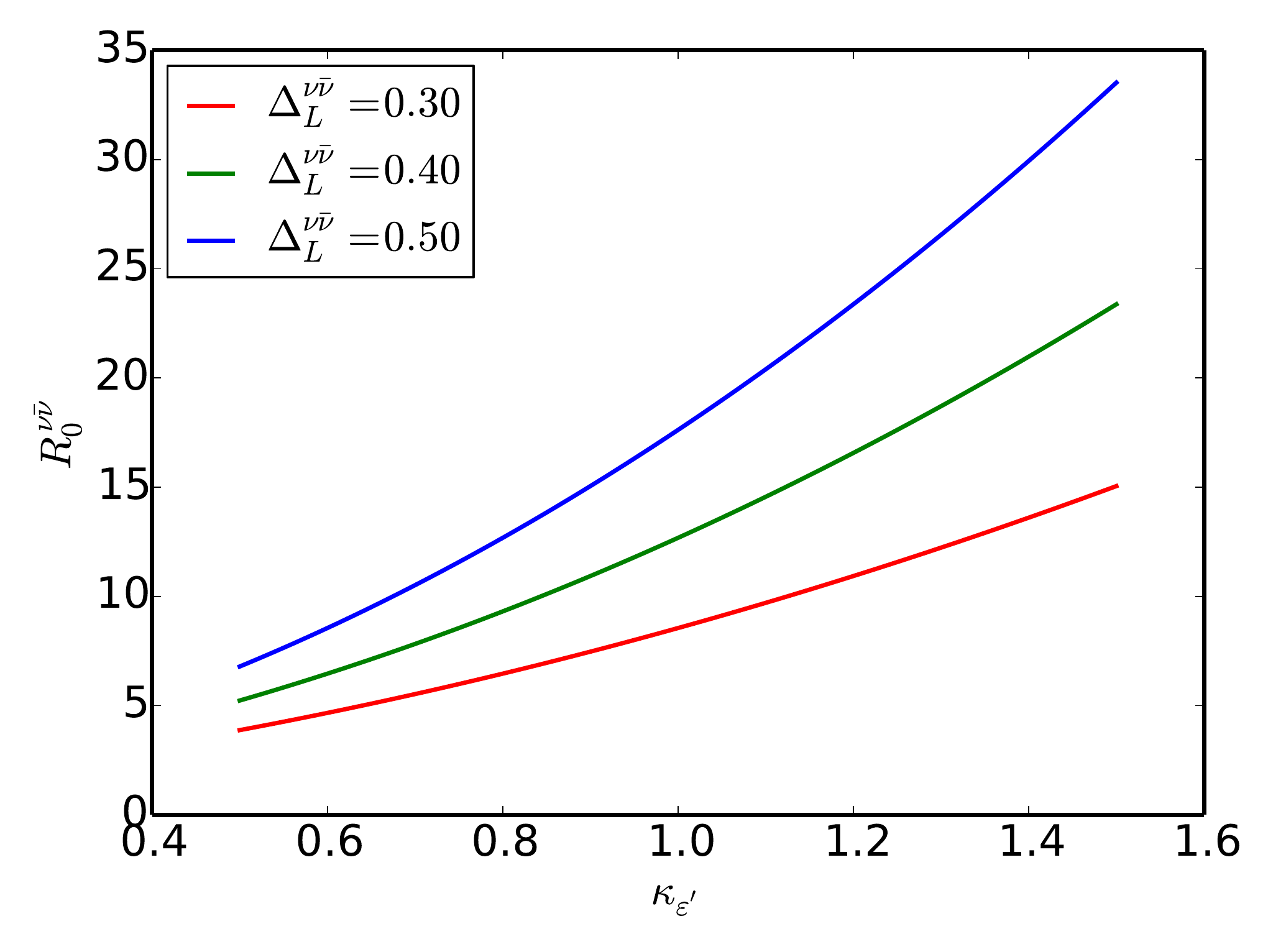}
\includegraphics[width = 0.45\textwidth]{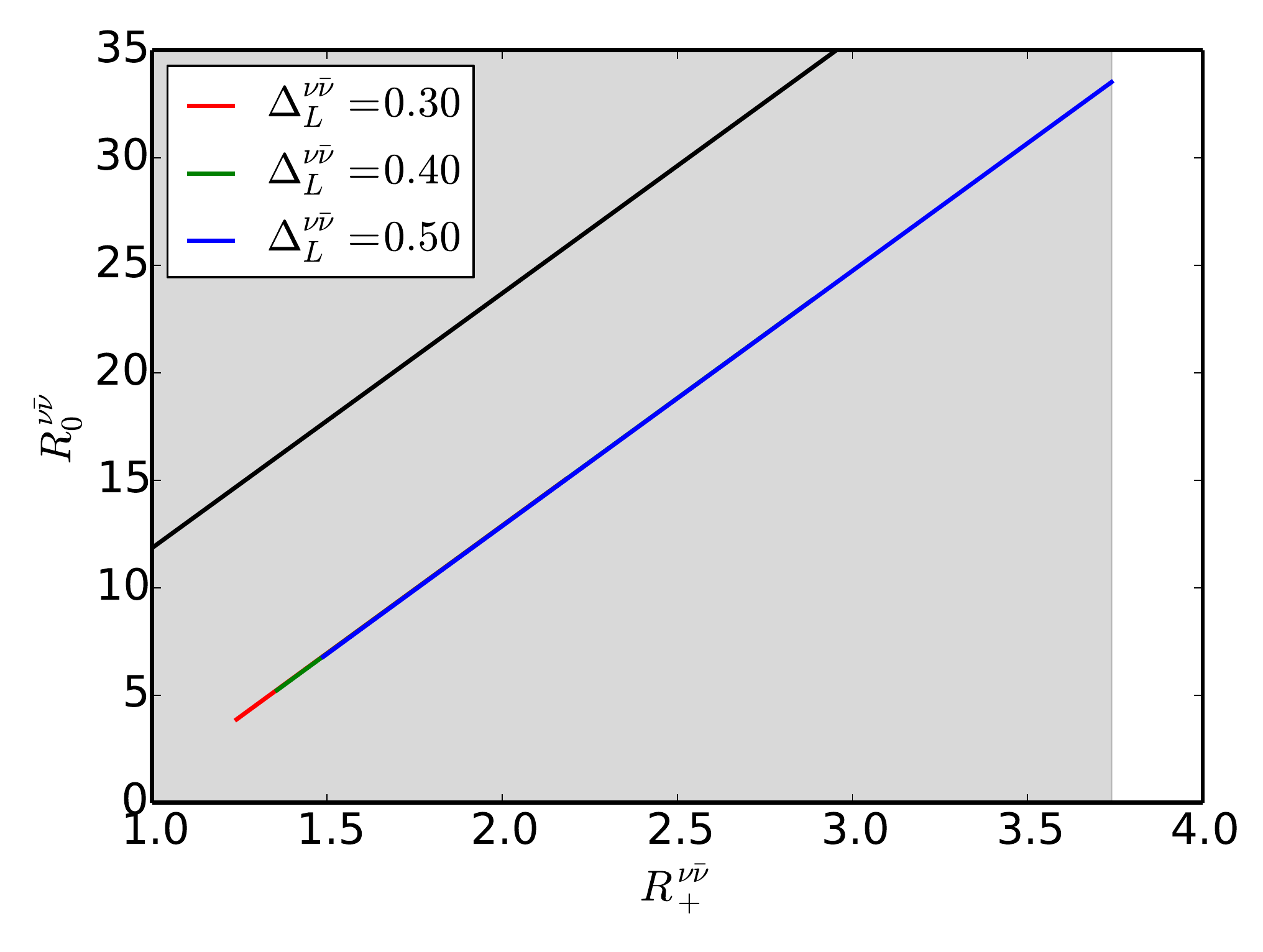}
\includegraphics[width = 0.45\textwidth]{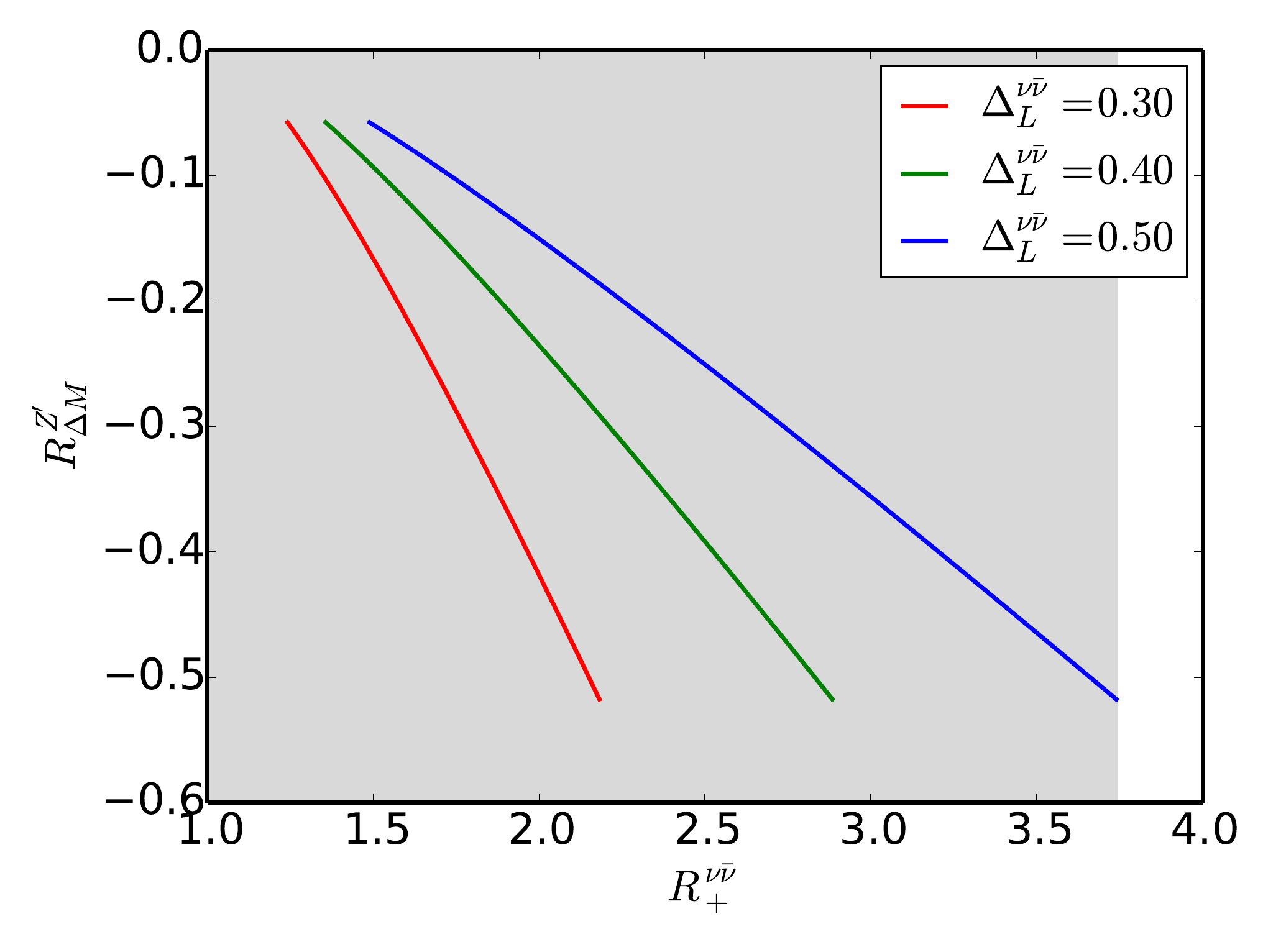}
\caption{ \it $R_+^{\nu\bar\nu}$ and  $R_0^{\nu\bar\nu}$, as functions 
of $\kepe$ for $\Delta^{\nu\bar\nu}_L(Z^\prime)=0.3,\,0.4,\, 0.5$ for {\rm QCDP} scenario. $M_{Z^\prime}=3\tev$. The dependence on $\keps$ is negligible. The upper 
black line in the lower left panel is the GN bound \cite{Grossman:1997sk}. In the fourth panel correlation of $R^{Z^\prime}_{\Delta M}$ with  $R_+^{\nu\bar\nu}$ is given. The experimental $1\sigma$ range for  $R_+^{\nu\bar\nu}$  is displayed by the grey band. From \cite{Buras:2015jaq}.
}\label{R5}~\\[-2mm]\hrule
\end{figure}

\begin{figure}[!tb]
 \centering
\includegraphics[width = 0.45\textwidth]{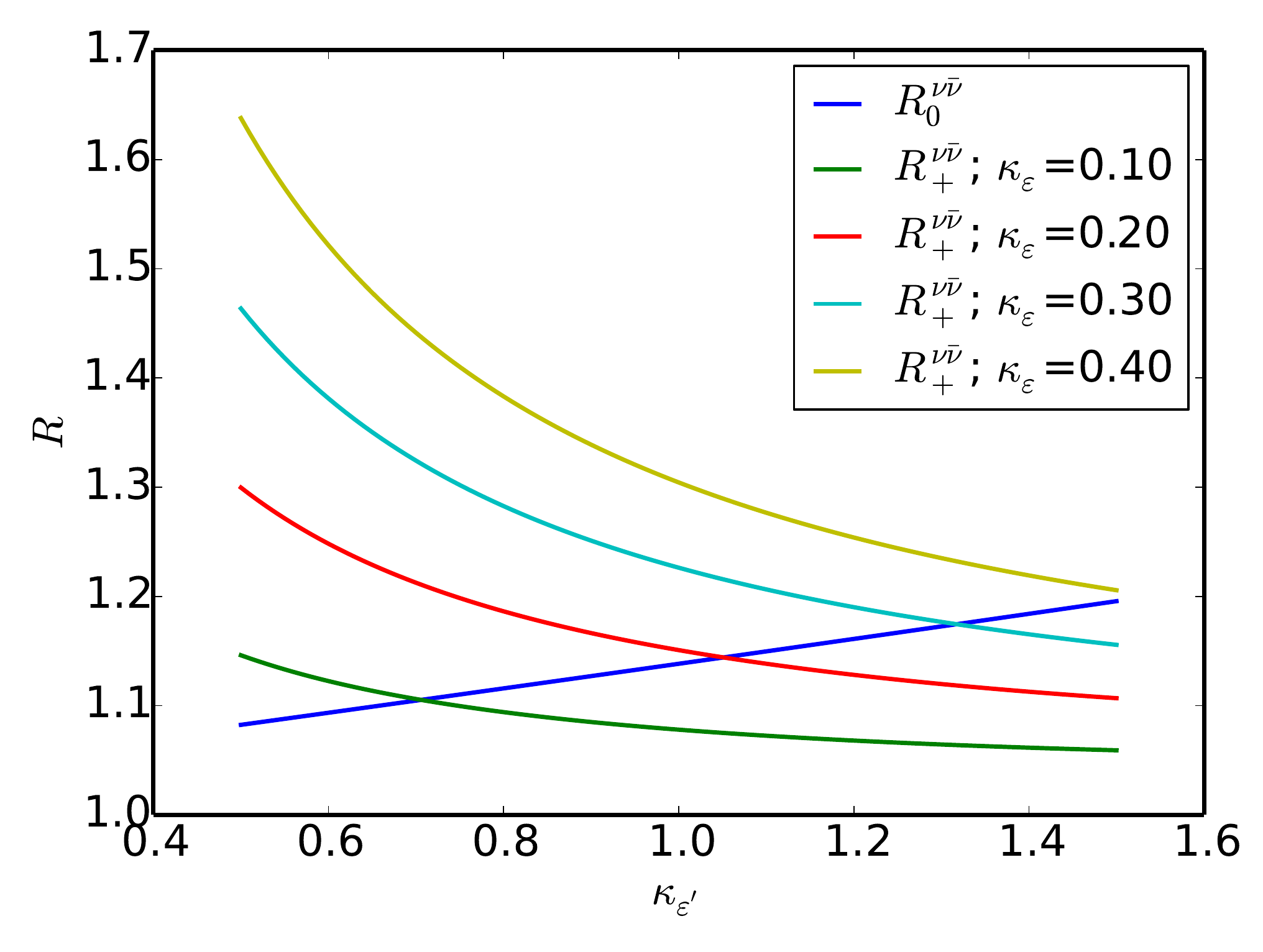}
\includegraphics[width = 0.45\textwidth]{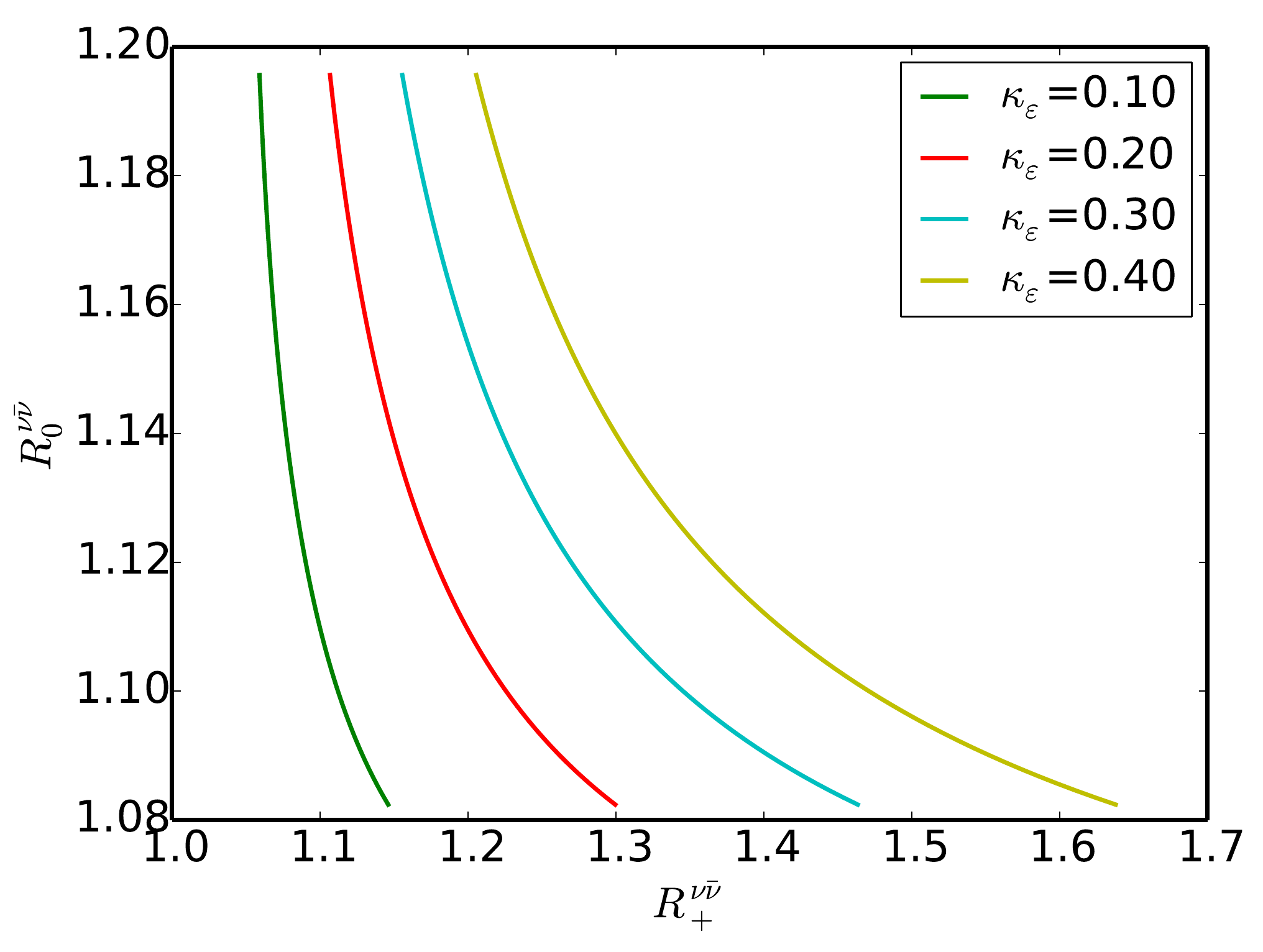}
\caption{ \it $R_0^{\nu\bar\nu}$ and  $R_+^{\nu\bar\nu}$, as functions 
of $\kepe$ for $\keps=0.1,\,0.2,\, 0.3,\, 0.4$ for EWP scenario. 
$\Delta^{\nu\bar\nu}_L(Z^\prime)=0.5$. From \cite{Buras:2015jaq}.
}\label{R6}~\\[-2mm]\hrule
\end{figure}

\subsection{Lessons on  NP  Patterns in $Z^\prime$ Scenarios}\label{LessonsZP}
$Z^\prime$ models 
exhibit quite different pattern of NP effects in the $K$ meson system than the 
LH and RH $Z$ scenarios. In $Z$ scenarios only electroweak 
penguin (EWP) $Q_8$ and 
$Q_8^\prime$ operators can contribute in an important manner because of flavour dependent diagonal $Z$ coupling to quarks. But in $Z^\prime$ models the diagonal quark couplings can be flavour universal so that QCD penguin operators (QCDP) ($Q_6,Q_6^\prime$) can dominate NP contributions to $\epe$. Interestingly,
the pattern of NP in rare $K$ decays 
depends on whether NP in $\epe$ is dominated by QCDP  or EWP operators.  Moreover, 
the striking difference from $Z$ scenarios, known already from previous 
studies, is the increased importance of the constraints from $\Delta F=2$ 
observables. This has two virtues in the presence of the $\epe$ constraint:
\begin{itemize}
\item
The real parts of the couplings are determined for not too a large $\keps$ from the $\varepsilon_K$ constraint even for LH and RH scenarios. 
\item
There is a large hierarchy between real and imaginary parts of the  flavour 
violating couplings implied by anomalies in  QCDP and EWP scenarios.
As shown in  \cite{Buras:2015jaq}
in the case of QCDP  imaginary parts dominate over the real ones, while
in the case of EWP this hierarchy is opposite  unless
the $\varepsilon_K$ anomaly is absent.
\end{itemize}

Because of a significant difference in the manner QCDP and EWP enter $\epe$, there are  striking differences in the implications 
for the correlation between $\kpn$ and $\klpn$ in these
two NP  scenarios if significant NP contributions to $\epe$ are required.

The plots in Figs.~\ref{R5} and \ref{R6} show clearly the differences between QCDP and EWP scenarios. We refer to  \cite{Buras:2015jaq} for more details, in 
particular analytic derivation of all these results. We extract from these 
results the following lessons:

{\bf Lesson 7:}
In the case of QCDP scenario the correlation between 
$\mathcal{B}(\klpn)$ and $\mathcal{B}(\kpn)$ takes place along the branch 
parallel to the GN bound. See lower left panel in Fig.~\ref{R5}. Moreover
this feature is independent of $M_{Z^\prime}$. Also the dependence on $\keps$ 
is negligible and we show therefore the dependence on $\Delta^{\nu\bar\nu}_L(Z^\prime)$.

{\bf Lesson 8:}
In the EWP scenario this correlation   between 
$\mathcal{B}(\klpn)$ and $\mathcal{B}(\kpn)$  
proceeds away from  this branch for diagonal quark couplings 
$\ord(1)$ if NP in $\varepsilon_K$ is present and it is very different 
from the one of the QCDP case as seen in  Fig.~\ref{R6}. As NP effects 
 turn out to be modest in this case we set $\Delta^{\nu\bar\nu}_L(Z^\prime)=0.5$. Only 
for the diagonal quark couplings $\ord(10^{-2})$ the requirement of 
shifting upwards $\epe$ implies large effects in $\kpn$ and $\klpn$ in EWP scenario. See  \cite{Buras:2015jaq}  for a detail discussion of this point.

{\bf Lesson 9:}
For fixed values of the neutrino and  diagonal quark couplings in $\epe$ the 
predicted enhancements of $\mathcal{B}(\klpn)$ and $\mathcal{B}(\kpn)$ 
are much larger when NP in QCDP is required to remove the 
$\epe$ anomaly than it is the case of EWP. This is simply related to the fact that the $\Delta I=1/2$ 
rule suppresses QCDP contributions to $\epe$ so that QCDP operators 
are less effective in enhancing $\epe$ than EWP operators and 
consequently the imaginary parts of the flavour violating $Z^\prime$ couplings 
are required to be larger. 

{\bf Lesson 10:}
 In QCDP scenario  $\Delta M_K$ is {\it suppressed} and this 
effect increases with increasing  $M_{Z^\prime}$ whereas in the EWP scenario 
 $\Delta M_K$ is {\it enhanced} and this effect decreases with increasing
 $M_{Z^\prime}$ as long as real couplings dominate.  Already on the basis of this property one could differentiate between 
these two scenarios when the SM prediction for $\Delta M_K$ improves.

\section{331 Flavour News}\label{sec:331}
The 331 models are based on the gauge group $SU(3)_C\times SU(3)_L\times U(1)_X$ \cite{Pisano:1991ee,Frampton:1992wt}.
In these models new contributions to $\epe$ and other flavour observables are  dominated by tree-level exchanges of a $Z^\prime$ with non-negligible contributions from tree-level $Z$ exchanges generated through the $Z-Z^\prime$ mixing. The size of these NP effects depends not only on $M_{Z^\prime}$ but in particular on a parameter $\beta$, which distinguishes between various 331 
models, on fermion representations under the gauge group and a
parameter $\tan\bar\beta$ present in the $Z-Z^\prime$ mixing. Extensive recent
analyses in these models can be found in  \cite{Buras:2012dp,Buras:2013dea,Buras:2014yna,Buras:2015kwd,Buras:2016dxz}. References to earlier analysis of flavour physics in 331 models can be found there and in  \cite{Diaz:2004fs,CarcamoHernandez:2005ka}.

\begin{table}[!tb]
{\renewcommand{\arraystretch}{1.3}
\begin{center}
\begin{tabular}{|c||c|c|c||c||c|c|c||c||c|c|c|}
\hline
MI  &     {\rm scen.} &  $\beta$ & $\tan {\bar \beta}$ & MI  & {\rm scen.} &  $\beta$ & $\tan {\bar \beta}$ &MI  & {\rm scen.} &  $\beta$ & $\tan {\bar \beta}$\\
\hline
M1 & $F_1$ & $-2/\sqrt{3}$ & 1 & M9 & $F_2$ & $-2/\sqrt{3}$ & 1 & M17 & $F_1$ & $-2/\sqrt{3}$ & 0.2
\\
M2 & $F_1$ & $-2/\sqrt{3}$ & 5 & M10 & $F_2$ & $-2/\sqrt{3}$ & 5 & M18 & $F_2$ & $-2/\sqrt{3}$ & 0.2
\\
M3 & $F_1$ & $-1/\sqrt{3}$ & 1 & M11 & $F_2$ & $-1/\sqrt{3}$ & 1 & M19 & $F_1$ & $-1/\sqrt{3}$ & 0.2
\\
M4 & $F_1$ & $-1/\sqrt{3}$ & 5 & M12 & $F_2$ & $-1/\sqrt{3}$ & 5 & M20 & $F_2$ & $-1/\sqrt{3}$ & 0.2
\\
M5 & $F_1$ & $1/\sqrt{3}$ & 1 & M13 & $F_2$ & $1/\sqrt{3}$ & 1 & M21 & $F_1$ & $1/\sqrt{3}$ & 0.2
\\
M6 & $F_1$ & $1/\sqrt{3}$ & 5 & M14 & $F_2$ & $1/\sqrt{3}$ & 5 & M22 & $F_2$ & $1/\sqrt{3}$ & 0.2
\\
M7 & $F_1$ & $2/\sqrt{3}$ & 1 & M15 & $F_2$ & $2/\sqrt{3}$ & 1 & M23 & $F_1$ & $2/\sqrt{3}$ & 0.2
\\
M8 & $F_1$ & $2/\sqrt{3}$ & 5 & M16 & $F_2$ & $2/\sqrt{3}$ & 5 & M24 & $F_2$ & $2/\sqrt{3}$ & 0.2
\\
\hline
\end{tabular}
\end{center}}
\caption{\it Definition of the various 331 models. From  \cite{Buras:2014yna}.
\label{tab:331models}}~\\[-2mm]\hrule
\end{table}
%%%%%%%%%%%%%%%%%%%%%%%%%%%%%%%%%%%%%%%

A detailed analysis of 331 models with different values of 
$\beta$, $\tan\bar\beta$ for two fermion representations $F_1$ and $F_2$, with 
the third SM quark generation belonging respectively to an antitriplet and a triplet under 
the $SU(3)_L$, has been presented in \cite{Buras:2014yna}. They are collected in Table~\ref{tab:331models}.
Requiring that these 24 models perform
at least as well as the SM as far as electroweak tests are concerned, seven models have been selected 
for a more detailed study of FCNC processes. These are
\be\label{favoured}
{\rm M9}, \quad {\rm M8},\quad {\rm M6}, \quad {\rm M11}, \quad {\rm M3}, \quad {\rm M16}, \quad {\rm M14}, \qquad {(\rm favoured)}
\ee
with the first five performing better than the SM while the last two 
basically as the SM.

A recent updated 
analyses have been presented in \cite{Buras:2015kwd,Buras:2016dxz} and we 
summarize the main results of these two papers putting the emphasize 
on the last analysis in \cite{Buras:2016dxz} which could take into account
new lattice QCD results from Fermilab Lattice and MILC Collaborations \cite{Bazavov:2016nty}   on $B^0_{s,d}-\bar B^0_{s,d}$ hadronic matrix elements.

The new analyses in \cite{Buras:2015kwd,Buras:2016dxz} show that the 
impact of  a required  enhancement of $\epe$  on other flavour observables 
is significant. The one in  \cite{Buras:2016dxz} also shows that the results 
are rather sensitive to the value of $\vcb$ which has been illustrated there
by choosing two values: $\vcb=0.040$ and $\vcb=0.042$.

The main findings of \cite{Buras:2015kwd,Buras:2016dxz} for $M_{Z^\prime}=3\tev$ are as follows:
\begin{itemize}
\item
Among seven 331 models in (\ref{favoured}) singled out through electroweak precision study only three  (M8, M9, M16) can provide for both choices of $\vcb$, significant shift of $\epe$ but not larger than $6\times 10^{-4}$, that is $\kepe\le 0.6$.
\item
The tensions between $\Delta M_{s,d}$  and $\varepsilon_K$, discussed in 
Section~\ref{sec:3a},
can be removed in  these models (M8, M9, M16) for both values of $\vcb$.
\item
Two of them (M8 and M9) can simultaneously suppress $B_s\to\mu^+\mu^-$ by
at most $10\%$ and $20\%$ for  $\vcb=0.042$ and  $\vcb=0.040$, 
respectively. This can still bring the theory within $1\sigma$ range of
the combined result from CMS and LHCb and for  $\vcb=0.040$ one can even 
reach the present central experimental value of this rate. The most recent result from ATLAS \cite{Aaboud:2016ire}, while not accurate, appears to confirm this picture. On the other hand
the maximal 
shifts in the Wilson coefficient $C_9$  are 
$C_9^\text{NP}=-0.1$ and $C_9^\text{NP}=-0.2$ for these two $\vcb$ values, 
respectively. This is only a moderate shift and these models 
do not really help in the case of $B_d\to K^*\mu^+\mu^-$ anomalies that 
require shifts as high as $C_9^\text{NP}=-1.0$ \cite{ Altmannshofer:2014rta,Descotes-Genon:2015uva}.
\item
In M16 the situation is opposite. The rate for $B_s\to\mu^+\mu^-$ can be reduced for  $M_{Z^\prime}=3\tev$ for the two $\vcb$ values by at most $3\%$ and $10\%$, 
respectively but 
  with the corresponding values $C_9^\text{NP}=-0.3$ and $-0.5$ the anomaly  in $B_d\to K^*\mu^+\mu^-$ can be significantly reduced.
\item
The  maximal shifts in $\epe$ decrease fast with increasing  $M_{Z^\prime}$ in the case of $\vcb=0.042$ but are practically unchanged for 
 $M_{Z^\prime}=10\tev$ when $\vcb=0.040$  is used.
\item
On the other hand for higher values of  $M_{Z^\prime}$ the effects in $B_s\to\mu^+\mu^-$ and $B_d\to K^*\mu^+\mu^-$ are much smaller.
We recall that  NP effects in rare $K$ decays and $B\to K(K^*)\nu\bar\nu$ remain small in all 331 models even for  $M_{Z^\prime}$ of few TeV. This could be challenged by NA62, KOTO and Belle II experiments in this decade.
\end{itemize}

\begin{figure}[!tb]
 \centering
\includegraphics[width = 0.47\textwidth]{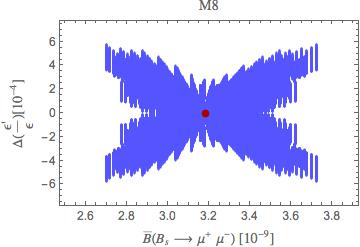}
\includegraphics[width = 0.47\textwidth]{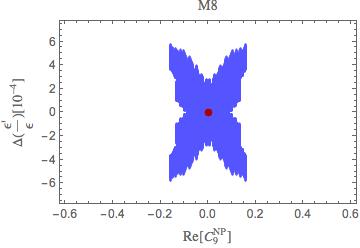}
\includegraphics[width = 0.47\textwidth]{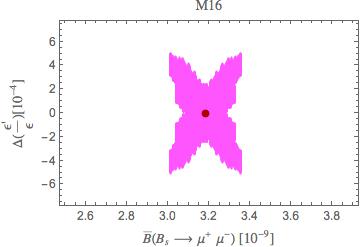}
\includegraphics[width = 0.47\textwidth]{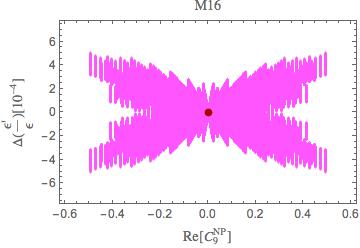}
\caption{ \it Correlations of $\Delta(\epe)$ with $B_s\to \mu^+\mu^-$ 
(left panels) and with $ C_9^{\rm NP}$ (right panels) for M8 and M16. 
 Red dots represent central SM values.
 $M_{Z^\prime}=3\tev$ and  $\vcb=0.040$.}
\label{CORR816vcb4}~\\[-2mm]\hrule
\end{figure}
We show these correlations for  $M_{Z^\prime}=3\tev$ and  $\vcb=0.040$ in Fig.~\ref{CORR816vcb4}.

All these results are valid for $\vub=0.0036$. For its inclusive value 
of $\vub=0.0042$, we find that for  $\vcb=0.040$ the maximal shifts in $\epe$ are increased to 
 $7.7 \times 10^{-4}$ and $8.8\times 10^{-4}$ for  $M_{Z^\prime}=3\tev$ and 
 $M_{Z^\prime}=10\tev$, respectively.

Thus the main message from \cite{Buras:2015kwd,Buras:2016dxz} is that NP 
contributions in 331 
models can simultaneously solve $\Delta F=2$ tensions, enhance $\epe$ and 
suppress either the rate for $B_s\to\mu^+\mu⁻$ or $C_9$ Wilson coefficient
without any significant effect on $\kpn$ and $\klpn$ and $b\to s\nu\bar\nu$
transitions. While sizable NP effects in $\Delta F=2$ observables and 
$\epe$  can persist for $M_{Z^\prime}$ outside the reach of the LHC, such 
effects in $B_s\to\mu^+\mu^-$ will only be detectable provided $Z^\prime$ 
will be discovered soon. 

\section{Outlook}
\subsection{2018 Visions}\label{Vision}
Let us begin the final section with a dream about the discovery of NP in 
$\kpn$  by the   NA62 experiment through
\be\label{NA62}
\mathcal{B}(\kpn)= (18.0\pm 2.0)\cdot 10^{-11},\qquad ({\rm NA62},~2018)\,.
\ee
It should be emphasized that such a result would be without any doubt 
a clear signal for NP. Moreover, looking at the grey bands in several figures shown by us, such a result would be truly tantalizing with a big impact on our field. Assuming then that the lattice values of $\bsi$ and $\bei$ will not be modified 
significantly and the $\epe$ anomaly will stay with us with $\kepe=1.0$ this 
measurement will allow to exclude certain scenarios and favour other ones.
But this will also depend on the allowed size of NP in $\varepsilon_K$, 
$\Delta M_K$ and rare $B_{s,d}$ decays.

\subsection{Open Questions}
There is no doubt that in the coming years $K$ meson physics will strike back, 
in particular through improved estimates of SM predictions for $\epe$, 
$\varepsilon_K$, $\Delta M_K$ and $K_L\to\mu^+\mu^-$ and through crucial 
measurements of the branching ratios for $\kpn$ and $\klpn$. Correlations 
with other meson systems, lepton flavour physics, electric dipole moments 
and other rare processes should allow us to identify NP at very short distance 
scales \cite{Buras:2013ooa} and we should hope that this physics will also be directly seen at the LHC. Let us then close this talk by listing most pressing questions in kaon flavour physics for the coming years. On the theoretical side we have:
\begin{itemize}
\item
{\bf What is the precise value of $\kepe$?} Here the answer will come not only from lattice QCD but also through improved values of the CKM parameters, NNLO QCD corrections and an improved understanding of FSI and isospin breaking effects. The NNLO QCD corrections should be available soon. 
 The recent analysis in the large $N$ approach in \cite{Buras:2016fys}
 indicates  that 
 FSI are likely to be important for the $\Delta I=1/2$  rule  in agreement with previous 
studies \cite{Antonelli:1995gw,Bertolini:1995tp,Pallante:1999qf,Pallante:2000hk,Buras:2000kx,Buchler:2001np,Buchler:2001nm,Pallante:2001he}, but much less relevant for $\epe$. But it is
important that other lattice groups beyond RBC-UKQCD collaboration make efforts to calculate 
 $\epe$ in the SM.
\item
{\bf What is the precise value of $\keps$?} Here the reduction of CKM uncertainties is most important. Also the large error in the charm contribution $\eta_{cc}$ should be decreased. In spite of these uncertainties the most recent analysis in \cite{Blanke:2016bhf} indicates that if no NP is  
present in $\varepsilon_K$, it is expected to be found in $\Delta M_{s,d}$.
\item
{\bf What is the precise value of $\Delta M_K$ in the SM?} The present calculations from  dual QCD approach \cite{Bijnens:1990mz,Buras:2014maa} and 
 lattice QCD \cite{Christ:2012se,Bai:2014cva} give values of  $\Delta M_K$ 
in the ballpark of its experimental value but with  uncertainties as high as 
$\pm 30\%$ which do not allow to conclude whether NP is required to contribute here  or not. Even the sign of possible NP contributions is unknown. Let us
hope this will be found out in this decade.
\item
{\bf What are the precise values of $\RE A_2$ and $\RE A_0$?} Again lattice QCD 
will play the crucial role here. The study of NP contributions to $\RE A_0$ 
can be found in \cite{Buras:2014sba}.
\end{itemize}

On the experimental side we have:
\begin{itemize}
\item
{\bf What is $\mathcal{B}(\kpn)$ from NA62?} We should know it in 2018.
\item
{\bf What is $\mathcal{B}(\klpn)$ from KOTO?} We should know it around the year 2020.
\item
{\bf Do $Z^\prime$ or other new particles with masses in the reach of 
the LHC exist?} We could know it already this year.
\end{itemize}

There are clearly other topics in kaon physics which we did not mention here.
In particular the study of $K^+ \rightarrow \pi^+ \ell^+ \ell^-$ and 
$K_L \rightarrow \pi^0 \ell^+ \ell^-$ will become more important when the 
theory improves. For recent analyses in the dual QCD approach and lattice QCD 
see \cite{Coluccio-Leskow:2016tsp} and \cite{Christ:2015aha}, respectively. 
Lattice QCD could also reduce the uncertainty in the charm contribution to 
$\kpn$ \cite{Christ:2016eae}. Another interesting issue is the violation 
of lepton flavour and lepton flavour universality in rare kaon decays \cite{Crivellin:2016vjc}.

I also expect that the interest in searching for NP behind the  
$\epe$ anomaly will increase. Beyond the papers discussed by us 
there is 
an interesting recent analysis in \cite{Kitahara:2016otd}, where it is shown 
that $\epe$ anomaly can be explained in the MSSM with squark masses above $3\tev$ while satisfying $\varepsilon_K$ constraint without fine-tuning of CP-phases or other parameters. More papers  will appear soon.

Definitely there are exciting times ahead of us!

\section*{ Acknowledgements}
 
I would like to thank Monika Blanke, Fulvia De Fazio and Jean-Marc G{\'e}rard  for exciting time we spent together 
analyzing the topics discussed in this talk.
I would like to thank the organizers of BEAUTY 2016  for inviting me to present these results.
The research presented in this report was dominantly financed and done in the context of the ERC Advanced Grant project ``FLAVOUR'' (267104).  It was also partially supported by the 
DFG cluster of excellence ``Origin and Structure of the Universe''.

\bibliographystyle{JHEP}
\bibliography{allrefs}
\end{document}